\documentclass[pra,aps,onecolumn,showpacs,epsf]{revtex4}
\usepackage{amssymb}
\usepackage{amsmath}
\usepackage{epsfig}
\usepackage{graphicx}
\usepackage{subfigure}
\usepackage[english]{babel}

\begin{document}

\title{Composite boson many-body theory for Frenkel excitons}
\author{Monique Combescot (1) and Walter Pogosov (1,2)}
\affiliation{(1) Institut des NanoSciences de Paris, Universite Pierre et Marie Curie,
CNRS, 140 rue de Lourmel, 75015 Paris}
\affiliation{(2) Institute for Theoretical and Applied Electrodynamics, Russian Academy
of Sciences, Izhorskaya 13/19, 125412 Moscow}
\date{\today }

\begin{abstract}
We present a many-body theory for Frenkel excitons which takes into account
their composite nature exactly. Our approach is based on four commutators
similar to the ones we previously proposed for Wannier excitons. They allow
us to calculate any physical quantity dealing with $N$ excitons in terms of
"Pauli scatterings" for carrier exchange in the absence of carrier
interaction and "interaction scatterings" for carrier interaction in the
absence of carrier exchange. We show that Frenkel excitons have a novel
"transfer assisted exchange scattering", specific to these excitons. It
comes from indirect Coulomb processes between localized atomic states. These
indirect processes, commonly called "electron-hole exchange" in the case of
Wannier excitons and most often neglected, are crucial for Frenkel excitons,
as they are the only ones responsible for the excitation transfer. We also
show that in spite of the fact that Frenkel excitons are made of electrons
and holes on the same atomic site, so that we could naively see them as
elementary particles, they definitely are composite objects, their composite
nature appearing through various properties, not always easy to guess. The
present many-body theory for Frenkel excitons is thus going to appear as
highly valuable to securely tackle their many-body physics, as in the case
of nonlinear optical effects in organic semiconductors.
\end{abstract}

\maketitle

\section{Introduction}

Even if nature makes things somewhat more complex, we can roughly say that
there are essentially two types of excitons: Wannier excitons \cite{Wannier}
and Frenkel excitons \cite{Frenkel}.

(i) Wannier excitons are found in conventional (inorganic) semiconductors.
They are constructed on valence and conduction extended states. Their Bohr
radius, controlled by Coulomb interaction between electrons and holes,
screened by the semiconductor dielectric constant, is as large as a few tens
of interatomic distances. This makes their binding energy rather small.

(ii) Frenkel excitons are the excitons of organic semiconductors. They are
constructed on highly localized atomic states. Their "Bohr radius",
controlled by the atomic wave functions, is of the order of a single
molecular block. This makes their binding energy as large as a fraction of
e.V. or more.

Wannier excitons made of two fermions, the conduction electron and valence
hole, were up to now commonly considered as elementary bosons, the possible
fermion exchanges being included through effective scatterings, by lack of
appropriate procedure. Over the last few years, we have developed this
missing many-body theory \cite{3,4}, through a procedure which allows us to
treat the composite nature of Wannier excitons exactly. This theory now has
its specific diagrammatic representation in terms of "Shiva diagrams" \cite%
{Monique-Odile} which not only allow to visualize the tricky $N$-body
exchanges which exist in these systems but also to calculate them readily.
This new many-body theory has been successfully applied to a variety of
physical effects such as Faraday rotation \cite{a} and oscillation \cite{a1}%
, spin precession \cite{b} and teleportation \cite{c}, polariton and exciton
Bose-Einstein condensation \cite{d}. In each of these effects, the composite
nature of the excitons plays a key role; in particular this new theory
allows us to show that the replacement of Wannier excitons by elementary
bosons misses the correct detuning behavior of all nonlinear optical effects
induced by nonabsorbed photons.

Since organic materials are of potential importance for new electronic
devices, the proper handling of many-body effects with Frenkel excitons
along a theory similar to the one we have developed for Wannier excitons, is
of high current interest. For readers not aware of this new composite
exciton many-body theory, its fundamental ideas are briefly explained in a
short review paper \cite{3}. More details can be found in an extended
Physics Report \cite{4} and the references it contains to more than 30
published works.

In order to tackle Frenkel excitons on a firm basis we have, in a previous
paper \cite{paper1}, rederived the basic concept leading to these excitons,
starting from first principles, i.e., a microscopic Hamiltonian written in
first quantization, with electron and hole kinetic energies, electron-ion,
electron-electron, and ion-ion Coulomb interactions (This ion-ion term,
which does not affect the electron motion, makes the system at hand neutral,
in this way insuring the convergence of the overall Coulomb contribution in
the large sample limit). Through a grouping of terms appropriate to highly
localized electron states - this grouping being different from the one
leading to valence and conduction bands on which the Wannier excitons are
constructed - we make appear the atomic states of each ion. When overlaps
between relevant atomic states of different ion sites are negligible, these
states can be used as a one-body basis to rewrite the system Hamiltonian in
second quantization. In a second step, we transform the ground and excited
atomic creation operators into electron and hole creation operators. Among
the various Coulomb terms which keep the number of electron-hole pairs
constant, one is insuring the local neutrality of the excitations. The
resulting set of lowest excited states, made of these locally neutral
excitations, with one electron and one hole on the same site, are degenerate
in the absence of Coulomb interaction between sites. We then introduce these
intersite interactions. Among them, one is a transfer term which destroys
one electron-hole pair on one site and recreates it on another site. This
transfer term splits the degenerate subspace made of one-site excitations
into linear combinations of these excitations which are nothing but the
Frenkel exciton states.

Frenkel excitons, made of electron-hole pairs localized on all possible
atomic sites, are composite bosons, definitely. Their many-body effects thus
have to be handled with extreme care if we want to fully trust the obtained
results. This is why it can appear as appropriate to construct for them a
full-proof procedure similar to the one we have used for Wannier excitons 
\cite{3,4,7,8,9}, these Wannier excitons also being linear combinations of
electrons and holes but in extended states - not in localized states as for
Frenkel excitons. Actually, due to this high localization, we can be tempted
to believe that carrier exchanges between Frenkel excitons should reduce to
zero; so that these excitons should behave as elementary bosons. In the
present work, we show that this idea is actually a very na\"{\i}ve one:
Frenkel excitons do behave as elementary bosons for some quantities, but
differ from them in other important respects, making their many-body physics
not easy to guess. This is why the many-body theory we here propose is going
to be highly valuable because it allows us to tackle many-body effects with
Frenkel excitons in a fully secure way.

Before going further, let us point out one important difference between
Wannier and Frenkel excitons: Wannier excitons are controlled by one length
for both, the interaction and the extension of the relative motion wave
function. This length is the Bohr radius induced by the screened Coulomb
interaction between electron and hole and calculated with the effective
masses of the conduction and valence bands. The situation for Frenkel
excitons is far more complex. We could at first think that Frenkel excitons
have two lengths: the atomic wave function extension for the electron-hole
pair "relative motion" and the Bohr radius for Coulomb interaction.
Actually, Frenkel excitons have not one but two relevant Coulomb
interactions: one for direct Coulomb processes between sites; the other for
indirect processes, this last interaction being the one responsible for the
excitation transfer between sites. In addition, as the electrons and holes
on which the Frenkel excitons are constructed are not free, the proper
effective mass to use in a "Bohr radius" is a priori unclear. Finally, the
tight-binding approximation underlying Frenkel exciton corresponds to reduce
the atomic wave function extension to zero, so that the length associated to
the electron-hole "relative motion" de facto disappears. Due to these
intrinsic difficulties and the expected interplay between these various
"physical lengths", the many-body physics of Frenkel excitons is going to be
far more subtle than the one of Wannier excitons, controlled by one length
only.

The goal of the present paper is to settle the basis for a composite boson
many-body theory appropriate to these Frenkel excitons. Using the formalism
presented in the present work, we are going, in the near future, to address
to the calculation of various observables, such as the ground state energy
of $N$ Frenkel excitons, the scattering rate of such excitons, and the
nonlinear susceptibility of materials with Frenkel excitons, the later
problem being very important for applications. Possible extension of the
present theory can be to include effects of interaction with the vibrational
modes, e.g., disorder and exciton-phonon coupling. This can be done by
adding the corresponding potential to the Frenkel exciton Hamiltonian.

The paper is organized as follows.

In Section II, we recall the main steps leading to Frenkel excitons, as
derived in our first work on these excitons \cite{paper1}.

In Section III, we derive the "Pauli scatterings" for carrier exchange
between two Frenkel excitons, starting from the "deviation-from-boson
operator" of these excitons \cite{7,8}.

In Section IV, we introduce Shiva diagrams for carrier exchanges and show
their similarities with Shiva diagrams for Wannier excitons \cite%
{Monique-Odile}. We also point out an intrinsic difficulty of these Frenkel
exciton Shiva diagrams linked to the tight-binding approximation which makes
atomic electron-hole states used to represent Frenkel excitons, not a clean
complete basis for all electron-hole pair states.

In Section V, we use these Pauli scatterings to calculate some relevant
scalar products, including the normalization factor $N!F_{N}$ for $N$
identical Frenkel excitons. We find that, as for Wannier excitons \cite%
{7,Monique-Tanguy}, $F_{N}$ which plays a key role in all exciton many-body
effects, decreases exponentially with $N$ due to carrier exchanges between
composite bosons. This $F_{N}$ turns out to have an extremely simple compact
form due to the fact that Frenkel excitons are characterized by one quantum
number only, the center-of-mass momentum $\mathbf{Q}$, while Wannier
excitons in addition have a relative motion index $\nu $. This is going to
greatly simplify all results on many-body effects linked to the composite
nature of these particles. We also discuss the physical meaning of the
relevant dimensionless parameter $\eta $\ which controls carrier exchanges
between excitons. We show that, in spite of our first understanding in the
case of Wannier excitons, this key parameter for many-body effects is not
linked to the spatial extension of the electron-hole relative motion wave
function but to the total number of excitons the sample can accommodate -
otherwise $\eta $ would reduce to zero in the case of Frenkel excitons. 
\textit{This new physical understanding of }$\eta $\textit{\ is one of the
important results of the present work; it provides a significant step toward
mastering the many-body physics of composite bosons, in general.}

In Section VI, we derive the closure relation which exists for $N$ exciton
states and we show that, while the prefactor for Wannier excitons \cite{11}
is $\left( 1/N!\right) ^{2}$, the prefactor for Frenkel excitons is $1/N!$
as for $N$ elementary bosons. \textit{\ }The fact that the prefactor for
closure relation of composite bosons made of 2 fermions is not always $%
(1/N!)^{2}$, as we found for Wannier excitons, is also one of the important
results of the paper. More on this prefactor can be found in Ref. \cite%
{Monique-Marc-Andre}.

In Section VII, we calculate the creation potential of one Frenkel exciton,
as necessary to properly describe the Coulomb interaction of this exciton
with the rest of the system \cite{8,9}. This Frenkel exciton creation
potential actually splits into three conceptually different terms: One term
comes from direct and exchange Coulomb processes inside the same site
responsible for local neutrality; the second term originates from the
indirect Coulomb processes between sites responsible for the excitation
transfer; the third term comes from direct Coulomb processes between sites,
this last contribution being similar to the one found for Wannier excitons.

In Section VIII, we calculate the interaction scatterings associated to the
three parts of the creation potential. We find that the one associated to
direct Coulomb processes between sites depends on the momentum transfer
between the "in" and "out" excitons, as in the case of Wannier excitons \cite%
{8,9}. The scattering associated to the neutrality term is physically very
strange as it does not depend on the momenta of the "in" and "out" excitons
involved in the scattering. It however disappears from the physics of these
excitons, when summation over all possible processes is performed. The third
scattering, associated to the excitation transfer between sites, can also be
seen as strange as it does not depend on the momentum transfer but just on
the momenta of the "out" excitons. This is physically linked to the fact
that the excitation transfer does not induce any momentum change. This third
scattering actually corresponds to a "transfer assisted exchange": the
coupling between two excitons comes from carrier exchange through a Pauli
scattering while the energy-like contribution to this scattering comes from
the transfer part of the Frenkel exciton energy resulting from indirect
Coulomb processes. It can be of interest to note that this transfer assisted
exchange scattering, somehow unexpected at first, has similarity with the
"photon assisted exchange scattering" between two polaritons we recently
identified \cite{12}. The symmetry between "out" and "in" states is actually
restored when considering energy conserving process, the scatterings being
then equal, as shown in Section IX. In this Section VIII, we also calculate
the Coulomb exchange scatterings when Coulomb interaction is followed by
carrier exchange. We show that, as a consequence of the tight-binding
approximation, the Coulomb exchange scatterings associated to direct and
transfer processes reduce to zero.

To enlighten the interplay between the Pauli scattering and the various
interaction scatterings between two Frenkel excitons, Section IX is devoted
as an exercise to the calculation of the matrix element of the Frenkel
exciton Hamiltonian in the two-exciton subspace. We make clear the part
coming from non-interacting elementary bosons, the part coming from
non-interacting composite excitons and the part coming from interaction
between these excitons. We also compare this matrix element with the one
obtained by assuming that Frenkel excitons are interacting elementary
bosons. We again show that, in order to have equality between matrix
elements for composite and bosonized Frenkel excitons, we are forced to use
physically unacceptable effective scatterings, i.e., scatterings which would
induce non-hermiticity in the Hamiltonian.

In Section X, we discuss the various results obtained on Frenkel excitons by
comparing them with similar results obtained for elementary bosons and for
Wannier excitons. We show that Frenkel excitons have similarities with both.
This in particular means that they definitely differ from elementary bosons
by various means which makes all intuitive handling of these excitons rather
dangerous.

Section XI is a brief state-of-the-art. We discuss the approaches proposed
by the two main groups who have tackled interaction between Frenkel excitons 
\cite{13,14,AgrGal,Agranovich,AgrTos,Davydov,CherMuk,Mukamel} and point out
some of the advantages of the present theory.

In the last Section, we conclude with a brief summary of the main results
derived in the present work.

This paper is definitely quite long: We have chosen to include all
fundamental results on the composite boson many-body theory for Frenkel
excitons in a unique paper, for the reader to have at hand all the necessary
tools to securely tackle any physical effect dealing with these excitons.
Results on the Pauli and interaction scatterings \cite{8,9}, the
normalization factor \cite{7,Monique-Tanguy}, the closure relation \cite{11}
and the Shiva diagram expansion \cite{Monique-Odile} were the purposes of
different papers at the time we were constructing a similar composite boson
many-body theory for Wannier excitons. Through all these works, we have
however learned what should be done to be fast and efficient; This is why we
can now propose a rather compact presentation for all these important pieces
of the composite Frenkel exciton puzzle.

\section{All the way to Frenkel excitons}

\subsection{Electron and hole creation operators}

Let us call $a_{n}^{\dagger }=a_{\nu =1,\text{ }n}^{\dagger }$ the operator
which creates one electron in the atomic excited state $\nu =1$ on site $n$
with $n$ running from 1 to the total number $N_{s}$ of sites in the periodic
lattice. In the same way, the hole creation operator $b_{n}^{\dagger
}=a_{\nu =0,\text{ }n}$ destroys one electron in the atomic ground state $%
\nu =0$ on site $n$. In the tight-binding limit in which the products of
atomic wave functions are such that $\varphi _{\nu ^{\prime }}^{\ast }(%
\mathbf{r-R}_{n^{\prime }})\varphi _{\nu }(\mathbf{r-R}_{n})\simeq 0$ for
all $n^{\prime }\neq n$ when $(\nu ,$ $\nu ^{\prime })=(0,$ $1)$, these
operators obey 
\begin{equation}
a_{n^{\prime }}a_{n}^{\dagger }+a_{n}^{\dagger }a_{n^{\prime }}=\left\{
a_{n^{\prime }},a_{n}^{\dagger }\right\} \simeq \delta _{n^{\prime }n}\simeq
\left\{ b_{n^{\prime }},b_{n}^{\dagger }\right\}  \tag{2.1}  \label{2.1}
\end{equation}%
\begin{equation}
\left\{ a_{n^{\prime }},a_{n}\right\} =\left\{ b_{n^{\prime }},b_{n}\right\}
=\left\{ a_{n^{\prime }},b_{n}^{\dagger }\right\} =0\simeq \left\{
a_{n^{\prime }},b_{n}\right\}  \tag{2.2}  \label{2.2}
\end{equation}%
They can thus be used as a one-body basis for electrons and holes in
problems in which only the ground and first excited atomic states play a
role.

\subsection{Electron-hole Hamiltonian}

The Hamiltonian for these free electrons and holes reads%
\begin{equation}
H_{eh}=\varepsilon _{e}\sum_{n}a_{n}^{\dagger }a_{n}+\varepsilon
_{h}\sum_{n}b_{n}^{\dagger }b_{n}  \tag{2.3}
\end{equation}%
where the electron and hole energies $\varepsilon _{e}$ and $\varepsilon
_{h} $ are slightly different from the excited and ground state atomic
energies due to Coulomb interactions with the "jellium" made of one electron
in the atomic ground state of each site $n$. A similar difference also
exists for Wannier excitons, the electron and hole energies being slightly
different from the conduction and valence electron energies.

The one-electron-hole-pair eigenstates of $H_{eh}$\ are $a_{n_{e}}^{\dagger
}b_{n_{h}}^{\dagger }\left\vert v\right\rangle $ (see Fig. 1(a)), their
energy being $\varepsilon _{e}+\varepsilon _{h}$. Since $n_{e}$ and $n_{h}$
can run from 1 to $N_{s}$, these states form a $N_{s}^{2}$ degenerate
subspace.

\subsection{Pair Hamiltonian}

If we now take into account the \textit{intrasite} direct and exchange
contribution of the Coulomb potential, the resulting pair Hamiltonian is
given by%
\begin{equation*}
H_{pair}=H_{eh}+V_{intra}
\end{equation*}%
\begin{equation}
V_{intra}=-\varepsilon _{0}\sum_{n}a_{n}^{\dagger }b_{n}^{\dagger
}b_{n}a_{n}=-\varepsilon _{0}\sum_{n}B_{n}^{\dagger }B_{n}  \tag{2.4}
\end{equation}%
where $B_{n}^{\dagger }=a_{n}^{\dagger }b_{n}^{\dagger }$ creates one
electron-hole pair on site $n$. The energy $\varepsilon _{0}$ gained by such
a pair reads as

\begin{equation}
\varepsilon _{0}=V_{\mathbf{R=0}}\left( _{0\ 0}^{1\ 1}\right) -V_{\mathbf{R=0%
}}\left( _{1\ 0}^{0\ 1}\right)  \tag{2.5}
\end{equation}%
The Coulomb couplings between atomic states $V_{\mathbf{R}}\left( _{\nu
_{1}^{\prime }\ \nu _{1}}^{\nu _{2}^{\prime }\ \nu _{2}}\right) $ correspond
to atom on site $n_{1}$, going from $\nu _{1}$ to $\nu _{1}^{\prime }$,
while atom on site $n_{2}$ goes from $\nu _{2}$ to $\nu _{2}^{\prime }$, the
two sites being at $\mathbf{R}_{n}=\mathbf{R}_{n_{1}}-\mathbf{R}_{n_{2}}$.
It is related to the ground and first excited atomic wave functions $\varphi
_{\nu }(\mathbf{r})$\ through%
\begin{equation}
V_{\mathbf{R}}\left( _{\nu _{1}^{\prime }\ \nu _{1}}^{\nu _{2}^{\prime }\
\nu _{2}}\right) =\int d\mathbf{r}_{1}d\mathbf{r}_{2}\varphi _{\nu
_{1}^{\prime }}^{\ast }(\mathbf{r}_{1})\varphi _{\nu _{2}^{\prime }}^{\ast }(%
\mathbf{r}_{2})\frac{e^{2}}{\left\vert \mathbf{r}_{1}-\mathbf{r}_{2}+\mathbf{%
R}\right\vert }\varphi _{\nu _{2}}(\mathbf{r}_{2})\varphi _{\nu _{1}}(%
\mathbf{r}_{1})  \tag{2.6}
\end{equation}%
Note that Eq. (2.6) makes the two terms of $\varepsilon _{0}$ real while the
orthogonality of atomic states makes it positive; so that $V_{intra}$ forces
local neutrality in the electron-hole excitation - as physically expected
due to the electrostatic energy cost to separate electron from hole.

Consequently, among the $N_{s}^{2}$ one-pair states $a_{n_{e}}^{\dagger
}b_{n_{h}}^{\dagger }\left\vert v\right\rangle $, the $N_{s}$ states with
the electron and hole on the same site have the same lowest energy (see Fig.
1 (b)). This leads to%
\begin{equation}
(H_{pair}-E_{pair})\left\vert n\right\rangle =0  \tag{2.7}
\end{equation}%
\begin{equation}
\left\vert n\right\rangle =a_{n}^{\dagger }b_{n}^{\dagger }\left\vert
v\right\rangle =B_{n}^{\dagger }\left\vert v\right\rangle  \tag{2.8}
\end{equation}%
where $E_{pair}=\varepsilon _{e}+\varepsilon _{h}-\varepsilon _{0}$. Note
that, since $a_{n}^{2}=0$, due to Eq. (2.2), the electron-hole operators $%
B_{n}$ are such that%
\begin{equation}
B_{n}^{2}=0  \tag{2.9}
\end{equation}

\subsection{Free exciton Hamiltonian}

The next step is to introduce the "electron-hole exchange" part of the
Coulomb interaction, i.e., the so-called "transfer term". It describes 
\textit{intersite indirect} Coulomb processes between $\nu =0$ and $\nu =1$
states and allows excitation transfer between sites (see Fig. 1(c)). This
leads to the "free exciton Hamiltonian" defined as%
\begin{equation*}
H_{X}^{(0)}=H_{pair}+V_{transf}
\end{equation*}%
\begin{equation*}
V_{transf}=\sum_{n_{1}\neq n_{2}}V_{\mathbf{R}_{n_{1}}-\mathbf{R}%
_{n_{2}}}\left( _{1\ 0}^{0\ 1}\right) a_{n_{1}}^{\dagger }b_{n_{1}}^{\dagger
}b_{n_{2}}a_{n_{2}}
\end{equation*}%
\begin{equation}
=\sum_{n_{1}\neq n_{2}}V_{\mathbf{R}_{n_{1}}-\mathbf{R}_{n_{2}}}\left( _{1\
0}^{0\ 1}\right) B_{n_{1}}^{\dagger }B_{n_{2}}  \tag{2.10}
\end{equation}%
The prefactor of this transfer term corresponds to Coulomb process in which
one electron-hole pair on site $n_{2}$ recombines, i.e., the atom on site $%
n_{2}$ is deexcited from $\nu =1$\ to $\nu =0$, while the atom on site $n_{1}
$\ is excited from $\nu =0$\ to $\nu =1$. This transfer term splits the $%
N_{s}$ degenerate subspace made of states $\left\vert n\right\rangle $ into $%
N_{s}$ Frenkel exciton states defined as%
\begin{equation}
\left\vert X_{\mathbf{Q}}\right\rangle =B_{\mathbf{Q}}^{\dagger }\left\vert
v\right\rangle =\frac{1}{\sqrt{N_{s}}}\sum_{n=1}^{N_{s}}e^{i\mathbf{Q.R}%
_{n}}\left\vert n\right\rangle   \tag{2.11}
\end{equation}%
While $\left\vert n\right\rangle $ corresponds to a localized excitation,
Frenkel excitons $\left\vert X_{\mathbf{Q}}\right\rangle $ correspond to
delocalized excitations through the sum over the $\mathbf{R}_{n}$ sites it
contains. By noting that%
\begin{equation*}
V_{transf}\left\vert X_{\mathbf{Q}}\right\rangle =\frac{1}{\sqrt{N_{s}}}%
\sum_{n\neq m}\sum V_{\mathbf{R}_{n}-\mathbf{R}_{m}}\left( _{1\ 0}^{0\
1}\right) e^{i\mathbf{Q}.\mathbf{R}_{m}}a_{n}^{\dagger }b_{n}^{\dagger
}\left\vert v\right\rangle 
\end{equation*}%
\begin{equation}
=\frac{1}{\sqrt{N_{s}}}\sum_{n}e^{i\mathbf{Q}.\mathbf{R}_{n}}a_{n}^{\dagger
}b_{n}^{\dagger }\left\vert v\right\rangle \sum_{m\neq n}e^{-i\mathbf{Q}.%
\mathbf{(\mathbf{R}}_{n}-\mathbf{\mathbf{R}}_{m}\mathbf{)}}V_{\mathbf{%
\mathbf{R}}_{n}-\mathbf{\mathbf{R}}_{m}}\left( _{1\ 0}^{0\ 1}\right)  
\tag{2.12}
\end{equation}%
while the sum over $m$ is independent on $n$ due to translational
invariance, we readily see that the $\left\vert X_{\mathbf{Q}}\right\rangle $
states are such that%
\begin{equation}
(H_{X}^{(0)}-E_{\mathbf{Q}})\left\vert X_{\mathbf{Q}}\right\rangle =0 
\tag{2.13}
\end{equation}%
where the exciton energy $E_{\mathbf{Q}}$ is related to the transfer
coupling through%
\begin{equation}
E_{\mathbf{Q}}=\varepsilon _{e}+\varepsilon _{h}-\varepsilon _{0}+\mathcal{V}%
_{\mathbf{Q}}  \tag{2.14}
\end{equation}%
\begin{equation}
\mathcal{V}_{\mathbf{Q}}=\sum_{\mathbf{R}\neq 0}e^{-i\mathbf{Q.R}}V_{\mathbf{%
R}}\left( _{1\ 0}^{0\ 1}\right) =\mathcal{V}_{\mathbf{Q}}^{\ast }  \tag{2.15}
\end{equation}%
Note that $\mathcal{V}_{\mathbf{Q}}$ is real as expected for an energy
since, due to Eq. (2.6), we do have%
\begin{equation}
V_{\mathbf{R}}^{\ast }\left( _{1\ 0}^{0\ 1}\right) =V_{\mathbf{R}}\left(
_{0\ 1}^{1\ 0}\right) =V_{-\mathbf{R}}\left( _{1\ 0}^{0\ 1}\right)  
\tag{2.16}
\end{equation}%
Also note that there is a plus sign in the exponential of the Frenkel
exciton creation operator while there is a minus sign in the exponential of
the Frenkel exciton energy.

While Wannier excitons are characterized by a center of mass momentum $%
\mathbf{Q}$ and a relative motion index, Frenkel excitons are characterized
by $\mathbf{Q}$ only. This is going to make the many-body physics of Frenkel
excitons far simpler than the one of Wannier excitons, the "in" excitons ($%
\mathbf{Q}_{1}$, $\mathbf{Q}_{2}$) and the "out" excitons ($\mathbf{Q}%
_{1}^{\prime }$, $\mathbf{Q}_{2}^{\prime }$) of a scattering process being
simply linked by $\mathbf{Q}_{1}^{\prime }+\mathbf{Q}_{2}^{\prime }=\mathbf{Q%
}_{1}+\mathbf{Q}_{2}$, due to momentum conservation, without any additional
degree of freedom.

By noting that the creation of an exciton $\mathbf{Q}$ and the creation of
an excitation on site $n$ are linked by%
\begin{equation}
B_{\mathbf{Q}}^{\dagger }=\frac{1}{\sqrt{N_{s}}}\sum_{n=1}^{N_{s}}e^{i%
\mathbf{Q.R}_{n}}B_{n}^{\dagger }  \tag{2.17}
\end{equation}%
\begin{equation}
B_{n}^{\dagger }=a_{n}^{\dagger }b_{n}^{\dagger }=\frac{1}{\sqrt{N_{s}}}%
\sum_{\mathbf{Q}}e^{-i\mathbf{Q.R}_{n}}B_{\mathbf{Q}}^{\dagger }  \tag{2.18}
\end{equation}%
while due to the lattice periodicity, we do have%
\begin{equation}
\sum_{\mathbf{Q}}e^{i\mathbf{Q.(\mathbf{R}}_{n^{\prime }}\mathbf{-\mathbf{%
\mathbf{R}}}_{n}\mathbf{)}}=N_{s}\delta _{n^{\prime }n}  \tag{2.19}
\end{equation}%
\begin{equation}
\sum_{n}e^{i\mathbf{(Q}^{\prime }-\mathbf{Q).\mathbf{R}}_{n}}=N_{s}\delta _{%
\mathbf{Q}^{\prime }\text{ }\mathbf{Q}}  \tag{2.20}
\end{equation}%
it is easy to show that the neutrality term $V_{intra}$\ in Eq. (2.4)\ reads
as $-\varepsilon _{0}\sum_{\mathbf{Q}}B_{\mathbf{Q}}^{\dagger }B_{\mathbf{Q}}
$ while the transfer term $V_{trans}\ $reads as $\sum_{\mathbf{Q}}\mathcal{V}%
_{\mathbf{Q}}B_{\mathbf{Q}}^{\dagger }B_{\mathbf{Q}}$. Consequently, the
free exciton Hamiltonian can be written as%
\begin{equation}
H_{X}^{(0)}=\varepsilon _{e}\sum_{n}a_{n}^{\dagger }a_{n}+\varepsilon
_{h}\sum_{n}b_{n}^{\dagger }b_{n}+S_{X}  \tag{2.21}
\end{equation}
\begin{equation}
S_{X}=\sum_{\mathbf{Q}}\zeta _{\mathbf{Q}}B_{\mathbf{Q}}^{\dagger }B_{%
\mathbf{Q}}  \tag{2.22}
\end{equation}%
where we have set 
\begin{equation}
\zeta _{\mathbf{Q}}=\mathcal{V}_{\mathbf{Q}}-\varepsilon _{0}  \tag{2.23}
\end{equation}%
Note that, while exciton operators can be used to rewrite the Coulomb part $%
S_{X}$ of this free exciton Hamiltonian, this is not possible for the
kinetic energy parts which still reads in terms of fermion operators. Also
note that, although coming from two-body Coulomb interaction, $S_{X}$
appears as a \textit{diagonal} operator in the Frenkel exciton subspace.
This is going to have important consequence for the scattering of two
Frenkel excitons associated to this part of the Hamiltonian.

\subsection{Interacting exciton Hamiltonian}

Coulomb interaction also contains \textit{intersite direct} contributions
coming from electron-electron and hole-hole Coulomb interactions, as well as
terms coming from direct electron-hole Coulomb processes, the exchange
electron-hole term between sites being already included in $H_{X}^{(0)}$
through $V_{transf}$. Consequently, the interacting exciton Hamiltonian
ultimately reads as%
\begin{equation}
H_{X}=H_{X}^{(0)}+V_{coul}  \tag{2.24}
\end{equation}%
\begin{equation}
V_{coul}=V_{ee}+V_{hh}+V_{eh}^{(dir)}  \tag{2.25}
\end{equation}%
where the three terms of the Coulomb interaction between electrons and
holes, shown in Fig. 1 (d, e, f), correspond to direct processes. They
precisely read%
\begin{equation}
V_{ee}=\frac{1}{2}\sum_{n_{1}\neq n_{2}}V_{R_{n_{1}}-R_{n_{2}}}\left( _{1\
1}^{1\ 1}\right) a_{n_{1}}^{\dagger }a_{n_{2}}^{\dagger }a_{n_{2}}a_{n_{1}} 
\tag{2.26}
\end{equation}%
\begin{equation}
V_{hh}=\frac{1}{2}\sum_{n_{1}\neq n_{2}}V_{R_{n_{1}}-R_{n_{2}}}\left( _{0\
0}^{0\ 0}\right) b_{n_{1}}^{\dagger }b_{n_{2}}^{\dagger }b_{n_{2}}b_{n_{1}} 
\tag{2.27}
\end{equation}%
\begin{equation}
V_{eh}^{(dir)}=-\sum_{n_{1}\neq n_{2}}V_{R_{n_{1}}-R_{n_{2}}}\left( _{0\
0}^{1\ 1}\right) b_{n_{1}}^{\dagger }a_{n_{2}}^{\dagger }a_{n_{2}}b_{n_{1}} 
\tag{2.28}
\end{equation}%
Note that none of these three potentials can be exactly written in terms of
exciton operators $B_{\mathbf{Q}}^{\dagger }$. Also note that the potential $%
V_{coul}$ plays a role on states having more than one electron-hole pair
since it contains operators acting on different sites, while each site can
be occupied by one electron-hole pair only - in the absence of spin degrees
of freedom. The introduction of these spin degrees of freedom and the quite
subtle polarization effects they induce, are out of the scope of the present
paper.

\section{Carrier exchanges}

\subsection{Free pair operators}

In the previous section, we made appear the excitations of one electron-hole
pair on site $n$ through the creation operators $B_{n}^{\dagger
}=a_{n}^{\dagger }b_{n}^{\dagger }$ (see Eq. (2.8)). From Eqs. (2.1, 2), it
is easy to show that the commutators of these free pair operators are such
that%
\begin{equation}
B_{n^{\prime }}B_{n}-B_{n}B_{n^{\prime }}=\left[ B_{n^{\prime }},B_{n}\right]
=0  \tag{3.1}  \label{3.1}
\end{equation}%
\begin{equation}
\left[ B_{n^{\prime }},B_{n}^{\dagger }\right] _{-}=\delta _{n^{\prime
}n}-D_{n^{\prime }n}  \tag{3.2}  \label{3.2}
\end{equation}%
where the deviation-from-boson operator $D_{n^{\prime }n}$ for free pairs is
given by%
\begin{equation}
D_{n^{\prime }n}=\delta _{n^{\prime }n}(a_{n}^{\dagger }a_{n}+b_{n}^{\dagger
}b_{n})  \tag{3.3}  \label{3.3}
\end{equation}%
Its commutator with another pair thus reduces to 
\begin{equation}
\left[ D_{n^{\prime }n},B_{m}^{\dagger }\right] =2\delta _{n^{\prime
}n}\delta _{nm}B_{m}^{\dagger }  \tag{3.4}  \label{3.4}
\end{equation}%
This quite simple result physically comes from the fact that electrons and
holes have highly localized atomic wave functions; so that electron-hole
pairs can feel each other by the Pauli exclusion principle when they are on
the same site only.

\subsection{Pauli scatterings for Frenkel excitons}

Let us now turn to delocalized excitations, i.e., Frenkel excitons with
creation operators given in terms of free pair operators by Eq. (2.17). From
the commutators of these free pairs, we readily find%
\begin{equation}
\left[ B_{\mathbf{Q}^{\prime }},B_{\mathbf{Q}}\right] =0  \tag{3.5}
\end{equation}%
\ 
\begin{equation}
\left[ B_{\mathbf{Q}^{\prime }},B_{\mathbf{Q}}^{\dagger }\right] =\delta _{%
\mathbf{Q}^{\prime }\mathbf{Q}}-D_{\mathbf{Q}^{\prime }\mathbf{Q}}  \tag{3.6}
\end{equation}%
where the deviation-from-boson operator for Frenkel excitons is given by 
\begin{equation}
D_{\mathbf{Q}^{\prime }\mathbf{Q}}=\Delta _{\mathbf{Q-Q}^{\prime
}}^{(e)}+\Delta _{\mathbf{Q-Q}^{\prime }}^{(h)}  \tag{3.7}
\end{equation}%
in which we have set%
\begin{equation}
\Delta _{\mathbf{P}}^{(e)}=\frac{1}{N_{s}}\sum_{n}e^{i\mathbf{P.R}%
_{n}}a_{n}^{\dagger }a_{n}  \tag{3.8}
\end{equation}%
and similarly for $\Delta _{\mathbf{P}}^{(h)}$ with $a_{n}^{\dagger }a_{n}$\
replaced by $b_{n}^{\dagger }b_{n}$.

Using the expression of free pair operators $B_{n}^{\dagger }$ in terms of
Frenkel exciton operators given in Eq. (2.18) and the system periodicity
through Eqs. (2.19, 20), it is easy to show that%
\begin{equation}
\left[ D_{\mathbf{Q}^{\prime }\mathbf{Q}},B_{\mathbf{P}}^{\dagger }\right]
_{-}=\frac{2}{N_{s}}B_{\mathbf{P+Q-Q}^{\prime }}^{\dagger }  \tag{3.9}
\end{equation}%
If we now compare this equation with the standard definition of the Pauli
scattering, deduced from the one for Wannier excitons, namely%
\begin{equation}
\left[ D_{\mathbf{Q}^{\prime }\mathbf{Q}},B_{\mathbf{P}}^{\dagger }\right]
_{-}=\sum_{\mathbf{P}^{\prime }}\left\{ \lambda \left( _{\mathbf{P}^{\prime
}\ \mathbf{P}}^{\mathbf{Q}^{\prime }\ \mathbf{Q}}\right) +\lambda \left( _{%
\mathbf{Q}^{\prime }\ \mathbf{P}}^{\mathbf{P}^{\prime }\ \mathbf{Q}}\right)
\right\} B_{\mathbf{P}^{\prime }}^{\dagger }  \tag{3.10}
\end{equation}%
we are led to identify the Pauli scattering of two Frenkel excitons with%
\begin{equation}
\lambda \left( _{\mathbf{P}^{\prime }\ \mathbf{P}}^{\mathbf{Q}^{\prime }\ 
\mathbf{Q}}\right) =\lambda \left( _{\mathbf{Q}^{\prime }\ \mathbf{P}}^{%
\mathbf{P}^{\prime }\ \mathbf{Q}}\right) =\frac{1}{N_{s}}\delta _{\mathbf{P}%
^{\prime }+\mathbf{Q}^{\prime },\text{ }\mathbf{P}+\mathbf{Q}}  \tag{3.11}
\end{equation}

When compared to the Pauli scatterings of Wannier excitons \cite{8,9}, the
one for Frenkel excitons appears as structureless: It barely contains the
momentum conservation expected for any scattering. This extremely simple
form of Pauli scattering can be traced back to the lack of degrees of
freedom in the electron-hole pairs on which Frenkel excitons are
constructed. Indeed, electrons and holes are in the two lowest atomic states
by construction, i.e., the states for which the tight-binding approximation
is valid. These $\left\vert \nu =(0,\text{ }1),\text{ }n\right\rangle $
states do not form a complete set for all electron-hole pairs but can be
used as a basis for the Frenkel exciton relevant states. In contrast,
Wannier excitons are constructed on all the extended electron-hole states $%
\mathbf{k}_{e}$ and $\mathbf{k}_{h}$ which have the same center-of-mass
momentum $\mathbf{k}_{e}+\mathbf{k}_{h}=\mathbf{Q}$. In addition to the
center-of-mass momentum $\mathbf{Q}$, Wannier excitons thus have a momentum $%
\mathbf{p}$ for the relative motion of the electron and the hole in the
exciton. This momentum $\mathbf{p}$ is related to the electron and hole
momenta on which Wannier exciton is made, through $\mathbf{k}_{e}=\mathbf{p+}%
\alpha _{e}\mathbf{Q}$ and $\mathbf{k}_{h}=-\mathbf{p+}\alpha _{h}\mathbf{Q}$
where $\alpha _{e}=1-\alpha _{h}=m_{e}/(m_{e}+m_{h})$.

\subsection{Many-body effects coming from carrier exchanges}

As for Wannier excitons, in order to derive many-body effects induced by
carrier exchanges between any number of Frenkel excitons in an easy way, it
is necessary to construct commutators similar to the ones of Eqs. (3.6, 9),
but for $N$ excitons \cite{Monique-Tanguy,13}. They read%
\begin{equation}
\left[ D_{\mathbf{P}^{\prime }\mathbf{P}},B_{\mathbf{Q}}^{\dagger N}\right]
=2\frac{N}{N_{s}}B_{\mathbf{Q+P-P}^{\prime }}^{\dagger }B_{\mathbf{P}%
}^{\dagger N-1}  \tag{3.12}
\end{equation}%
\begin{equation*}
\left[ B_{\mathbf{P}^{\prime }},B_{\mathbf{Q}}^{\dagger N}\right] =NB_{%
\mathbf{Q}}^{\dagger N-1}\left( \delta _{\mathbf{P}^{\prime }\mathbf{Q}}-D_{%
\mathbf{P}^{\prime }\mathbf{Q}}\right) 
\end{equation*}%
\begin{equation}
-\frac{N(N-1)}{N_{s}}B_{\mathbf{Q}}^{\dagger N-2}B_{2\mathbf{Q-P}^{\prime
}}^{\dagger }  \tag{3.13}
\end{equation}%
as easy to recover by iteration. These commutators are just the ones for
Wannier excitons (see Eq. (5) in Ref. \cite{Monique-Tanguy}) with the Pauli
scattering replaced by its value for Frenkel excitons, namely $%
N_{s}^{-1}\delta _{\mathbf{Q}^{\prime }+\mathbf{P}^{\prime },\mathbf{Q}+%
\mathbf{P}}$.

\section{Shiva diagrams}

Shiva diagrams for carrier exchanges between Wannier excitons \cite%
{Monique-Odile} have been shown to be extremely powerful. They not only
allow to visualize all carrier exchanges which take place between excitons,
but also to calculate the physical effects these exchanges induce readily.
This makes them as valuable for the many-body physics of composite bosons as
the Feynman diagrams for the many-body physics of elementary quantum
particles. This is why it can be of interest to settle similar Shiva
diagrams for carrier exchanges between Frenkel excitons. However, as shown
below, their intuitive handling turns out to be less obvious due to the fact
that the $B_{n}^{\dagger }\left\vert v\right\rangle $ states on which
Frenkel excitons are constructed, do not form a clean complete basis for 
\textit{all} one-pair states - in contrast to the free conduction electron
and valence hole states on which the Wannier excitons are constructed.

\subsection{Pauli scattering for two Frenkel excitons}

Although possibly surprising at first, the very simple Pauli scattering of
Frenkel excitons found in Eq. (3.11) still has the same formal structure as
the one for Wannier excitons \cite{8,9}. Indeed, by writing the wave
function of one electron on site $n$ as $\left\langle r\right\vert
a_{n}^{\dag }\left\vert v\right\rangle =\varphi _{\nu =1}(\mathbf{r}-\mathbf{%
R}_{n})$ and the wave function of one hole on site $n$ as $\left\langle
r\right\vert b_{n}^{\dag }\left\vert v\right\rangle =\varphi _{\nu =0}^{\ast
}(\mathbf{r}-\mathbf{R}_{n})$, where $\varphi _{\nu =0,1}$ are the atomic
wave functions, the exciton $\mathbf{Q}$ wave function\ reads as%
\begin{equation}
\left\langle \mathbf{r}_{h},\mathbf{r}_{e}|\mathbf{Q}\right\rangle =\Phi _{%
\mathbf{Q}}(e,h)=\frac{1}{\sqrt{N_{s}}}\sum_{n}e^{i\mathbf{Q.R}_{n}}\varphi
_{e}(\mathbf{r}_{e}-\mathbf{R}_{n})\varphi _{h}(\mathbf{r}_{h}-\mathbf{R}%
_{n})  \tag{4.1}
\end{equation}%
where we have set $\varphi _{e}=\varphi _{\nu =1}$ and $\varphi _{h}=\varphi
_{\nu =0}^{\ast }$. If we now calculate the diagram corresponding to hole
exchange between two excitons shown in Fig. 2(a), it reads in terms of the
Frenkel exciton wave functions given in Eq. (4.1) as%
\begin{equation}
\int d\left\{ r\right\} \Phi _{\mathbf{P}^{\prime }}^{\ast
}(e_{1},h_{2})\Phi _{\mathbf{Q}^{\prime }}^{\ast }(e_{2},h_{1})\Phi _{%
\mathbf{Q}}(e_{2},h_{2})\Phi _{\mathbf{P}}(e_{1},h_{1})  \tag{4.2}
\end{equation}%
This quantity contains integrals like 
\begin{equation}
\int dr_{e}\varphi _{e}^{\ast }(\mathbf{r}_{e}-\mathbf{R}_{n^{\prime
}})\varphi _{e}(\mathbf{r}_{e}-\mathbf{R}_{n})\simeq \delta _{n^{\prime }n} 
\tag{4.3}
\end{equation}%
for highly localized atomic states; so that, due to Eq. (2.20), the value of
this fermion exchange diagram (4.2) reduces to%
\begin{equation}
\frac{1}{N_{s}^{2}}\sum_{n}e^{i(-\mathbf{P}^{\prime }-\mathbf{Q}^{\prime }+%
\mathbf{P}+\mathbf{Q).R}_{n}}=\frac{1}{N_{s}}\delta _{\mathbf{P}^{\prime }+%
\mathbf{Q}^{\prime },\text{ }\mathbf{P}+\mathbf{Q}}  \tag{4.4}
\end{equation}%
which is nothing but the value of the Pauli scattering $\lambda \left( _{%
\mathbf{P}^{\prime }\ \mathbf{P}}^{\mathbf{Q}^{\prime }\ \mathbf{Q}}\right) $
obtained in Eq. (3.11). Consequently, as for Wannier excitons, the Pauli
scattering of two Frenkel excitons can be represented by the Shiva diagram
for fermion exchange between two excitons, shown in Fig. 2(a). The unique
(relevant) difference comes from the fact that, due to the lack of relative
motion index, the Pauli scattering for Frenkel excitons is the same for an
electron exchange or a hole exchange.

\subsection{Carrier exchange between more than two excitons}

If we now turn to the carrier exchange between three Frenkel excitons which
corresponds to the Shiva diagram of Fig. 2(b) and we calculate it along the
standard rules for Shiva diagrams \cite{Monique-Odile}, i.e., by just
writing what we read on the diagram, we find using Eqs. (4.1, 3) and Eq.
(2.20) that%
\begin{equation*}
\lambda \left( 
\begin{array}{cc}
\mathbf{Q}_{3}^{\prime } & \mathbf{Q}_{3} \\ 
\mathbf{Q}_{2}^{\prime } & \mathbf{Q}_{2} \\ 
\mathbf{Q}_{1}^{\prime } & \mathbf{Q}_{1}%
\end{array}%
\right) =\int d\left\{ r\right\} \Phi _{\mathbf{Q}_{1}^{\prime }}^{\ast
}(e_{1},h_{2})\Phi _{\mathbf{Q}_{2}^{\prime }}^{\ast }(e_{3},h_{1})\Phi _{%
\mathbf{Q}_{3}^{\prime }}^{\ast }(e_{2},h_{3})
\end{equation*}%
\begin{equation*}
\Phi _{\mathbf{Q}_{1}}(e_{1},h_{1})\Phi _{\mathbf{Q}_{2}}(e_{2},h_{2})\Phi _{%
\mathbf{Q}_{3}}(e_{3},h_{3})
\end{equation*}%
\begin{equation*}
=\frac{1}{N_{s}^{2}}\delta _{\mathbf{Q}_{1}^{\prime }+\mathbf{Q}_{2}^{\prime
}+\mathbf{Q}_{3}^{\prime },\text{ }\mathbf{Q}_{1}+\mathbf{Q}_{2}+\mathbf{Q}%
_{3}}
\end{equation*}%
This result agrees with the decomposition of three-body exchange in terms of
a succession of two-body exchanges shown in Fig. 2(b), namely%
\begin{equation}
\lambda \left( 
\begin{array}{cc}
\mathbf{Q}_{3}^{\prime } & \mathbf{Q}_{3} \\ 
\mathbf{Q}_{2}^{\prime } & \mathbf{Q}_{2} \\ 
\mathbf{Q}_{1}^{\prime } & \mathbf{Q}_{1}%
\end{array}%
\right) =\sum_{\mathbf{P}}\lambda \left( _{\mathbf{Q}_{3}^{\prime }\ \ 
\mathbf{P}}^{\mathbf{Q}_{2}^{\prime }\ \mathbf{Q}_{3}}\right) \lambda \left(
_{\mathbf{Q}_{1}^{\prime }\ \ \mathbf{Q}_{1}}^{\mathbf{P}\ \ \ \ \mathbf{Q}%
_{2}}\right)   \tag{4.6}
\end{equation}

It is easy to check that this is also true for the four-body exchange shown
in Fig. 2(c). And so on... So that the Shiva diagram for carrier exchange
between $N$ excitons just corresponds to%
\begin{equation*}
\frac{1}{N_{s}^{N-1}}\delta _{\mathbf{Q}_{1}^{\prime }+\mathbf{\ldots }+%
\mathbf{Q}_{N}^{\prime },\text{ }\mathbf{Q}_{1}+\mathbf{\ldots }+\mathbf{Q}%
_{N}}
\end{equation*}

\subsection{Difficulties with Frenkel excitons}

In spite of the above results, such a very visual way to calculate Shiva
diagrams between $N$ excitons and their decomposition in terms of exchanges
between two excitons, encounter a major problem when dealing with the double
exchange shown in Fig. 2(d): Indeed, Frenkel excitons $\mathbf{Q}%
_{1}^{\prime }$ and $\mathbf{Q}_{1}$ are clearly made with the same
electron-hole pair (see Fig. 2(e)); so that we should find $\delta _{\mathbf{%
Q}_{1}^{\prime }\mathbf{Q}_{1}}\delta _{\mathbf{Q}_{2}^{\prime }\mathbf{Q}%
_{2}}$ as for Wannier excitons. In contrast, two Pauli scatterings between
two Frenkel excitons give%
\begin{equation*}
\sum_{\mathbf{P,P}^{\prime }}\lambda \left( _{\mathbf{Q}_{1}^{\prime }\ \ 
\mathbf{P}}^{\mathbf{Q}_{2}^{\prime }\ \ \mathbf{P}^{\prime }}\right)
\lambda \left( _{\mathbf{P}\ \ \ \mathbf{Q}_{1}}^{\mathbf{P}^{\prime }\ \ 
\mathbf{Q}_{2}}\right) =\frac{1}{N_{s}^{2}}\sum_{\mathbf{P,P}^{\prime
}}\delta _{\mathbf{Q}_{1}^{\prime }+\mathbf{Q}_{2}^{\prime },\mathbf{P}%
^{\prime }+\mathbf{P}}\delta _{\mathbf{P}^{\prime }+\mathbf{P},\mathbf{Q}%
_{1}+\mathbf{Q}_{2}}
\end{equation*}%
\begin{equation}
=\frac{1}{N_{s}}\delta _{\mathbf{Q}_{1}^{\prime }+\mathbf{Q}_{2}^{\prime },%
\mathbf{Q}_{1}+\mathbf{Q}_{2}}  \tag{4.7}
\end{equation}%
which is nothing but the Pauli scattering $\lambda \left( _{\mathbf{Q}%
_{1}^{\prime }\ \ \mathbf{Q}_{1}}^{\mathbf{Q}_{2}^{\prime }\ \ \mathbf{Q}%
_{2}}\right) $ and thus definitely different from an identity.

This problem must be associated to the fact that, while for Wannier
excitons, the two possible ways to couple two electrons and two holes into
two excitons leads to \cite{7} 
\begin{equation}
B_{n}^{\dagger }B_{j}^{\dagger }=-\sum_{m\text{ }n}\lambda \left( _{m\
i}^{n\ j}\right) B_{m}^{\dagger }B_{n}^{\dagger }  \tag{4.8}
\end{equation}%
such a rearrangement is not possible for Frenkel excitons; so that a similar
relation does not exist for them. Indeed, using Eq. (2.17), we do have%
\begin{equation}
B_{\mathbf{P}}^{\dagger }B_{\mathbf{Q}}^{\dagger }=\frac{1}{N_{s}}\sum_{n%
\text{ }n^{\prime }}e^{i\mathbf{P.R}_{n}}e^{i\mathbf{Q.R}_{n^{\prime
}}}a_{n}^{\dagger }b_{n}^{\dagger }a_{n^{\prime }}^{\dagger }b_{n^{\prime
}}^{\dagger }  \tag{4.9}
\end{equation}%
Due to the tight-binding approximation, we cannot make an exciton out of the
electron-hole operator $a_{n}^{\dagger }b_{n^{\prime }}^{\dagger }$ except
for $n^{\prime }=n$, the product $a_{n}^{\dagger }b_{n}^{\dagger
}a_{n^{\prime }}^{\dagger }b_{n^{\prime }}^{\dagger }$ then reducing to 0,
according to Eq. (2.9).

The difficulty to rewrite $B_{\mathbf{Q}}^{\dagger }B_{\mathbf{Q}^{\prime
}}^{\dagger }$ in terms of two other Frenkel excitons, when compared to
Wannier excitons, can be mathematically assigned to the fact that the free
electron-hole pairs $a_{k_{e}}^{\dagger }b_{k_{h}}^{\dagger }\left\vert
v\right\rangle $ form a clean complete basis for one electron and one hole
while this is not true for the trapped pairs $a_{n}^{\dagger }b_{n}^{\dagger
}\left\vert v\right\rangle $. As previously discussed, these states can be
used as a basis in the tight-binding limit only, i.e., when the two lowest
atomic (bound) states are the only relevant states of the problem. However,
Eq. (4.8) as well as the identity between diagrams 2(d)-(f) for Wannier
excitons, relies on an exact closure relation for all the \textit{%
intermediate} electron-hole states on which the sums are taken while this
closure relation using the two lowest atomic states is only approximate in
the case of Frenkel excitons.

This may lead us to think that, since there is one way only to construct two
Frenkel excitons out of two electron-hole pairs, many-body effects coming
from fermion exchanges between Frenkel excitons which result from their
composite nature, must be somehow suppressed. The situation is actually far
more subtle, as shown in the next sections.

\section{Calculation of some relevant scalar products}

As for Wannier excitons, physical effects dealing with Frenkel excitons
ultimately read in terms of scalar products of Frenkel exciton states. Let
us now calculate a few of them along the line we have used for Wannier
excitons \cite{7,13}. Due to the structureless form of the Pauli scattering,
these scalar products turn out to be far simpler than the ones for Wannier
excitons. This makes the Frenkel exciton many-body physics induced by
carrier exchanges also far simpler.

\subsection{Scalar product of two-exciton states}

Using Eqs. (3.2,4), the scalar product of two-free-pair states reads as%
\begin{equation*}
\left\langle v\right\vert B_{n_{1}^{\prime }}B_{n_{2}^{\prime
}}B_{n_{2}}^{\dag }B_{n_{1}}^{\dag }\left\vert v\right\rangle =\left\langle
v\right\vert B_{n_{1}^{\prime }}\left\{ B_{n_{2}}^{\dag }B_{n_{2}^{\prime
}}-\delta _{n_{2}^{\prime }\text{ }n_{2}}-D_{n_{2}^{\prime }n_{2}}\right\}
B_{n_{1}}^{\dag }\left\vert v\right\rangle 
\end{equation*}%
\begin{equation}
=\delta _{n_{1}^{\prime }\text{ }n_{1}}\delta _{n_{2}^{\prime }\text{ }%
n_{2}}+\delta _{n_{1}^{\prime }\text{ }n_{2}}\delta _{n_{2}^{\prime }\text{ }%
n_{1}}-2\delta _{n_{1}^{\prime }\text{ }n_{1}}\delta _{n_{2}^{\prime }\text{ 
}n_{2}}\delta _{n_{1}^{\prime }\text{ }n_{2}^{\prime }}  \tag{5.1}
\end{equation}%
So that $\left\langle v\right\vert B_{n}B_{n^{\prime }}B_{n^{\prime }}^{\dag
}B_{n}^{\dag }\left\vert v\right\rangle =1-\delta _{n^{\prime }n}$ reduces
to 0 when $n^{\prime }=n$ as expected since a given site cannot be occupied
by two electrons in the absence of degeneracy. This shows that while there
are $N_{s}$ one-pair states $B_{n}^{\dag }\left\vert v\right\rangle $, the
number of different two-pair states $B_{m}^{\dag }B_{n}^{\dag }\left\vert
v\right\rangle $ is $N_{s}(N_{s}-1)/2$ since $m$ must be different from $n$,
while $B_{n}^{\dag }B_{m}^{\dag }=B_{m}^{\dag }B_{n}^{\dag }$.

If we now turn to two-exciton states, the same procedure leads, using Eqs.
(3.6, 9), to%
\begin{equation}
\left\langle v\right\vert B_{\mathbf{Q}_{1}^{\prime }}B_{\mathbf{Q}%
_{2}^{\prime }}B_{\mathbf{Q}_{2}}^{\dag }B_{\mathbf{Q}_{1}}^{\dag
}\left\vert v\right\rangle =\delta _{\mathbf{Q}_{1}^{\prime }\text{ }\mathbf{%
Q}_{1}}\delta _{\mathbf{Q}_{2}^{\prime }\text{ }\mathbf{Q}_{2}}+\delta _{%
\mathbf{Q}_{1}^{\prime }\text{ }\mathbf{Q}_{2}}\delta _{\mathbf{Q}%
_{2}^{\prime }\text{ }\mathbf{Q}_{1}}-\frac{2}{N_{s}}\delta _{\mathbf{Q}%
_{1}^{\prime }+\mathbf{Q}_{2}^{\prime }\text{ }\mathbf{Q}_{1}+\mathbf{Q}_{2}}
\tag{5.2}
\end{equation}%
By noting that the last term of the above bracket is nothing but $-\lambda
\left( _{\mathbf{Q}_{1}^{\prime }\ \mathbf{Q}_{1}}^{\mathbf{Q}_{2}^{\prime
}\ \mathbf{Q}_{2}}\right) -\lambda \left( _{\mathbf{Q}_{2}^{\prime }\ 
\mathbf{Q}_{1}}^{\mathbf{Q}_{1}^{\prime }\ \mathbf{Q}_{2}}\right) $, we see
that this result which is shown in Fig. 3, is formally the same as the one
for Wannier excitons, namely \cite{7}%
\begin{equation}
\left\langle v\right\vert B_{m}B_{n}B_{i}^{\dag }B_{j}^{\dag }\left\vert
v\right\rangle =\delta _{mi}\delta _{nj}-\lambda \left( _{m\ i}^{n\
j}\right) +(m\leftrightarrow n)  \tag{5.3}
\end{equation}%
with ($m$, $n$, $i$, $j$) replaced by ($\mathbf{Q}_{1}^{\prime }$, $\mathbf{Q%
}_{2}^{\prime }$, $\mathbf{Q}_{1}$, $\mathbf{Q}_{2}$).

It follows from Eq. (5.2) that the norm of two Frenkel exciton states%
\begin{equation}
\left\langle v\right\vert B_{\mathbf{Q}_{1}}B_{\mathbf{Q}_{2}}B_{\mathbf{Q}%
_{2}}^{\dag }B_{\mathbf{Q}_{1}}^{\dag }\left\vert v\right\rangle =1+\delta _{%
\mathbf{Q}_{1}\text{ }\mathbf{Q}_{2}}-\frac{2}{N_{s}}  \tag{5.4}
\end{equation}%
differs from 0 for $\mathbf{Q}_{1}=\mathbf{Q}_{2}$; so that there are $%
N_{s}^{2}/2$ different two-exciton states $B_{\mathbf{Q}_{1}}^{\dag }B_{%
\mathbf{Q}_{2}}^{\dag }\left\vert v\right\rangle $ since $B_{\mathbf{Q}%
_{1}}^{\dag }B_{\mathbf{Q}_{2}}^{\dag }=B_{\mathbf{Q}_{2}}^{\dag }B_{\mathbf{%
Q}_{1}}^{\dag }$ while there only are $N_{s}(N_{s}-1)/2$ different two-pair
states $B_{m}^{\dag }B_{n}^{\dag }\left\vert v\right\rangle $. Consequently,
as for Wannier excitons, the two-Frenkel exciton states form an overcomplete
set \cite{11}.

Note that, while the overcompleteness of the two Wannier-exciton states
readily follows from Eq. (5.3), we have shown that such a relation - which
comes from the two ways to bind two free electrons and two free holes into
two Wannier excitons - does not exist for Frenkel excitons due to the
tight-binding approximation which is at the basis of the Frenkel exciton
concept. Nevertheless, two-Frenkel-exciton states, like two-Wannier-exciton
states, form an overcomplete set.

\subsection{Scalar product of three-exciton states}

If we now turn to the scalar product of three-exciton states shown in Fig.
4, we find using the commutators of Eqs. (3.6, 9)%
\begin{equation*}
\left\langle v\right\vert B_{\mathbf{Q}_{3}^{\prime }}B_{\mathbf{Q}%
_{2}^{\prime }}B_{\mathbf{Q}_{1}^{\prime }}B_{\mathbf{Q}_{1}}^{\dag }B_{%
\mathbf{Q}_{2}}^{\dag }B_{\mathbf{Q}_{3}}^{\dag }\left\vert v\right\rangle
=\left\{ \delta _{\mathbf{Q}_{1}^{\prime }\text{ }\mathbf{Q}_{1}}\delta _{%
\mathbf{Q}_{2}^{\prime }\text{ }\mathbf{Q}_{2}}\delta _{\mathbf{Q}%
_{3}^{\prime }\text{ }\mathbf{Q}_{3}}+5\text{ perm.}\right\} 
\end{equation*}%
\begin{equation*}
-\frac{2}{N_{s}}\left\{ \delta _{\mathbf{Q}_{1}^{\prime }+\mathbf{Q}%
_{2}^{\prime }\text{ }\mathbf{Q}_{1}+\mathbf{Q}_{2}}\delta _{\mathbf{Q}%
_{3}^{\prime }\text{ }\mathbf{Q}_{3}}+8\text{ perm.}\right\} 
\end{equation*}%
\begin{equation}
+\frac{12}{N_{s}^{2}}\delta _{\mathbf{Q}_{1}^{\prime }+\mathbf{Q}%
_{2}^{\prime }+\mathbf{Q}_{3}^{\prime }\text{ }\mathbf{Q}_{1}+\mathbf{Q}_{2}+%
\mathbf{Q}_{3}}  \tag{5.5}
\end{equation}

This result agrees with the standard rules \cite{Monique-Odile} for
calculating scalar products through Shiva diagrams shown in Fig. 4, with the
values of fermion exchanges between two and three Frenkel excitons, given in
Eqs. (4.2, 4) and Eq. (4.5).

\subsection{Scalar product of $N$ identical exciton states}

As for Wannier excitons, the normalization factor of $N$ identical Frenkel
excitons plays a crucial role in all physically relevant matrix elements
involving $N$ excitons, since most of them usually are in the same state.
Let us write this normalization factor as \cite{7,Monique-Tanguy}%
\begin{equation}
\left\langle v\right\vert B_{\mathbf{Q}}^{N}B_{\mathbf{Q}}^{\dag
N}\left\vert v\right\rangle =N!F_{N}  \tag{5.6}
\end{equation}%
while $N!$\ is the value this scalar product would have if Frenkel excitons
were elementary bosons. We are going to show that, in contrast to Wannier
excitons for which $F_{N}$ depends on the state ($\nu $, $\mathbf{Q}$) in
which the $N$ excitons are, $F_{N}$ for Frenkel excitons depends on $N$ but
not on the exciton state considered, labelled by $\mathbf{Q}$. This property
is going to greatly simplify all results on many-body effects with Frenkel
excitons.

(i) Pedestrian way to calculate $F_{N}$

The pedestrian calculation of $F_{N}$ relies on the commutator (3.10). While 
$\left\langle v\right\vert B_{\mathbf{Q}}B_{\mathbf{Q}}^{\dag }\left\vert
v\right\rangle =1$ so that $F_{1}=1$, it is easy to show that%
\begin{equation*}
\left\langle v\right\vert B_{\mathbf{Q}}^{2}B_{\mathbf{Q}}^{\dag
2}\left\vert v\right\rangle =2\left( 1-\frac{1}{N_{s}}\right) 
\end{equation*}%
\begin{equation}
\left\langle v\right\vert B_{\mathbf{Q}}^{3}B_{\mathbf{Q}}^{\dag
3}\left\vert v\right\rangle =3!\left( 1-\frac{1}{N_{s}}\right) \left( 1-%
\frac{2}{N_{s}}\right)   \tag{5.7}
\end{equation}%
and so on... This leads us to conclude that $F_{N}$ should read as 
\begin{equation}
F_{N}=\left( 1-\frac{1}{N_{s}}\right) ...\left( 1-\frac{N-1}{N_{s}}\right) =%
\frac{N_{s}!}{(N_{s}-N)!N_{s}^{N}}  \tag{5.8}
\end{equation}%
which makes $F_{N}$ independent of the Frenkel exciton $\mathbf{Q}$
considered, a result somewhat surprising at first.

(ii) More elaborate derivation

A more elaborate way to calculate $F_{N}$\ makes use of the recursion
relation between the $\left\langle v\right\vert B_{\mathbf{Q}}^{N}B_{\mathbf{%
Q}}^{\dag N}\left\vert v\right\rangle $ easy to obtain from Eq. (3. 13).
This equation leads to%
\begin{equation*}
\left\langle v\right\vert B_{\mathbf{Q}}^{N}B_{\mathbf{Q}}^{\dag
N}\left\vert v\right\rangle \equiv \left\langle v\right\vert B_{\mathbf{Q}%
}^{N-1}B_{\mathbf{Q}}B_{\mathbf{Q}}^{\dag N}\left\vert v\right\rangle 
\end{equation*}%
\begin{equation}
=\left( N-\frac{N(N-1)}{N_{s}}\right) \left\langle v\right\vert B_{\mathbf{Q}%
}^{N-1}B_{\mathbf{Q}}^{\dag N-1}\left\vert v\right\rangle   \tag{5.9}
\end{equation}%
So that%
\begin{equation}
F_{N}=\left( 1-\frac{N-1}{N_{s}}\right) F_{N-1}  \tag{5.10}
\end{equation}%
from which the expression of $F_{N}$ given in Eq. (5.8) readily follows.

(iii) Recursion relation between the $F_{N}$'s

Even if $F_{N}$ for Frenkel excitons has a nicely compact expression, it can
be of interest to note that, due to Eq. (5.10), $F_{N}$ can be written as a
sum of all the other $F_{N}$'s according to%
\begin{equation}
F_{N}=F_{N-1}-\frac{N-1}{N_{s}}F_{N-2}+\frac{(N-1)(N-2)}{N_{s}^{2}}%
F_{N-3}-\ldots  \tag{5.11}
\end{equation}

This series just is the analog of the one for Wannier excitons, namely (see
Eq. (15) Ref. \cite{Monique-Tanguy})%
\begin{equation}
F_{N}=\sum_{n=1}^{N}(-1)^{n-1}\frac{(N-1)!}{(N-n)!}\sigma _{n}F_{N-n} 
\tag{5.12}
\end{equation}%
where the $\sigma _{n}$'s are the Shiva diagrams for fermion exchanges with $%
n$ excitons $\mathbf{Q}$ on both sides. Indeed, as shown in Section IV,
these Shiva diagrams in the case of Frenkel excitons reduce to $%
1/N_{s}^{n-1}\delta _{\mathbf{Q}_{1}^{\prime }+\mathbf{\ldots }+\mathbf{Q}%
_{n}^{\prime },\text{ }\mathbf{Q}_{1}+\mathbf{\ldots }+\mathbf{Q}_{n}}$,
i.e., $1/N_{s}^{n-1}$\ when all the \ $\mathbf{Q}$'s are equal.

(iv) Density expansion

From Eq. (5.10), we find 
\begin{equation}
\frac{F_{N-1}}{F_{N}}=\frac{N_{s}}{N_{s}+1-N}\simeq 1+O\left( \frac{N}{N_{s}}%
\right)  \tag{5.13}
\end{equation}%
for $1\ll N\ll N_{s}$. This result is again fully consistent with $%
F_{N-1}/F_{N}=1+O(\eta )$ as found for Wannier excitons \cite{Monique-Tanguy}
where $\eta =N(a_{x}/L)^{D}$ with $N$ being the number of excitons at hand, $%
a_{x}$ the exciton Bohr radius or relative motion wave function extension, $%
L $ the sample size and $D$ the space dimension. Indeed, the ratio of the
sample volume $L^{D}$ divided by the Wannier exciton volume $a_{x}^{D}$ is
nothing but the maximum number of excitons the sample can accept in the
absence of spin degrees of freedom. This thus makes $\eta $ for Wannier
excitons equal to the ratio of the number of excitons at hand divided by the
maximum possible number of excitons, this ratio being small in the dilute
limit, i.e., when excitons exist. In the case of Frenkel excitons, it is
clear that the maximum number of excitons has nothing to do with the Frenkel
exciton relative motion extension because this extension reduces to zero in
the tight binding limit. This maximum exciton number is just the number $%
N_{s}$ of atomic sites; so that the small dimensionless parameter for these
excitons, associated to their density, must be identified with%
\begin{equation}
\eta =\frac{N}{N_{s}}  \tag{5.14}
\end{equation}

This expression leads us to reconsider the physical understanding of the
parameter which controls carrier exchanges between composite bosons. While
the spatial extension of the electron-hole relative motion wave function
seemed at first to us a relevant length for the control of carrier
exchanges, the expression of $\eta $ we here find for Frenkel excitons shows
that the physically relevant quantity for the low density expansion of
carrier exchanges cannot be this length - otherwise $\eta $ would reduce to
zero in the tight-binding limit - but must be the maximum number of excitons
we can put in the sample volume $V$. With this new understanding, it becomes
easy to grasp why carrier exchanges are not totally suppressed in Frenkel
exciton systems.

If we now come back to the exact value of $F_{N}$ given in Eq. (5.8), we can
show that, as for Wannier excitons, it decreases exponentially with $N\eta $%
. This is done by using the Stirling formula $p!\simeq p^{p}e^{-p}\sqrt{2\pi
p}$ in Eq. (5.8). For $N_{s}$ and $N_{s}-N$ both large, we find%
\begin{equation}
F_{N}\simeq e^{-N}(1-\eta )^{-N\left( \frac{1}{\eta }-1\right) }(1-\eta
)^{-1/2}  \tag{5.15}
\end{equation}

Since $a^{b}=e^{b\ln a}$, we then get when all the $N\eta $'s are small%
\begin{equation}
F_{N}\simeq e^{-N\left( \frac{\eta }{2}+\frac{\eta ^{2}}{6}+...\right) } 
\tag{5.16}
\end{equation}%
A similar exponential decrease is also found for Wannier excitons.

\section{Closure relation for Frenkel excitons}

We have shown that, due to the existence of a clean closure relation for the
electrons and holes on which the Wannier excitons are constructed, these
composite boson excitons have a nicely compact closure relation which reads 
\cite{11}%
\begin{equation}
I=\frac{1}{\left( N!\right) ^{2}}\sum B_{i_{1}}^{\dagger }\ldots
B_{i_{N}}^{\dagger }\left\vert v\right\rangle \left\langle v\right\vert
B_{i_{N}}\ldots B_{i_{1}}  \tag{6.1}
\end{equation}%
the closure relation for elementary bosons being just the same with $%
1/\left( N!\right) ^{2}$ replaced by $1/N!$.

We are going to show that the situation for Frenkel excitons is rather
different due to the tight-binding approximation underlying the Frenkel
exciton concept.

\subsection{One electron-hole pair / one Frenkel exciton}

Starting from the closure relation for electrons in atomic site $n$%
\begin{equation}
I=\sum a_{n}^{\dagger }\left\vert v\right\rangle \left\langle v\right\vert
a_{n}  \tag{6.2}
\end{equation}%
and the similar relation for holes, we can construct the closure relation
for one-electron-hole pairs. It reads%
\begin{equation*}
I=\sum_{n\text{ }n^{\prime }}\sum a_{n}^{\dagger }b_{n^{\prime }}^{\dagger
}\left\vert v\right\rangle \left\langle v\right\vert b_{n^{\prime }}a_{n}
\end{equation*}%
\begin{equation}
=\sum_{n}B_{n}^{\dagger }\left\vert v\right\rangle \left\langle v\right\vert
B_{n}+\sum_{n\neq n^{\prime }}\sum a_{n}^{\dagger }b_{n^{\prime }}^{\dagger
}\left\vert v\right\rangle \left\langle v\right\vert b_{n^{\prime }}a_{n} 
\tag{6.3}
\end{equation}%
Using Eqs. (2.18, 20), it is easy to show that the first sum can be
rewritten as%
\begin{equation*}
\sum_{\mathbf{Q}^{\prime }\text{ }\mathbf{Q}}B_{\mathbf{Q}^{\prime
}}^{\dagger }\left\vert v\right\rangle \left\langle v\right\vert B_{\mathbf{Q%
}}\text{ }\frac{1}{N_{s}}\sum_{n}e^{i(\mathbf{Q}^{\prime }-\mathbf{Q})%
\mathbf{.R}_{n}}=\sum_{\mathbf{Q}}B_{\mathbf{Q}}^{\dagger }\left\vert
v\right\rangle \left\langle v\right\vert B_{\mathbf{Q}}
\end{equation*}%
while the second sum of Eq. (6.3) can be dropped out since it corresponds to
projections over states way off by an energy $\varepsilon _{0}$, their
electrons and holes being in different sites. Consequently, the closure
relation for one-Frenkel exciton states can be approximated by%
\begin{equation}
I\simeq \sum_{\mathbf{Q}}B_{\mathbf{Q}}^{\dagger }\left\vert v\right\rangle
\left\langle v\right\vert B_{\mathbf{Q}}  \tag{6.4}
\end{equation}

\subsection{Two free pairs / two Frenkel excitons}

Let us now turn to two-pairs. The closure relation for two electrons reads%
\begin{equation}
I=\frac{1}{2!}\sum a_{n_{1}}^{\dagger }a_{n_{2}}^{\dagger }\left\vert
v\right\rangle \left\langle v\right\vert a_{n_{2}}a_{n_{1}}  \tag{6.5}
\end{equation}%
and similarly for holes. By rewriting this hole closure relation as%
\begin{equation*}
I=\frac{1}{2!}\sum_{n_{2}^{\prime }}\left\{ \sum_{n_{1}^{\prime
}=(n_{1},n_{2})}b_{n_{1}^{\prime }}^{\dagger }b_{n_{2}^{\prime }}^{\dagger
}\left\vert v\right\rangle \left\langle v\right\vert b_{n_{2}^{\prime
}}b_{n_{1}^{\prime }}\right. 
\end{equation*}%
\begin{equation*}
\left. +\sum_{n_{1}^{\prime }\neq (n_{1},n_{2})}b_{n_{1}^{\prime }}^{\dagger
}b_{n_{2}^{\prime }}^{\dagger }\left\vert v\right\rangle \left\langle
v\right\vert b_{n_{2}^{\prime }}b_{n_{1}^{\prime }}\right\} 
\end{equation*}%
and by splitting the sum over $n_{2}^{\prime }$ in the same way, we find
that the closure relation for two electron-hole pairs restricted to the
degenerate subspace having an energy $(\varepsilon _{e}+\varepsilon
_{h}-\varepsilon _{0})$ can be written as%
\begin{equation*}
I\simeq \left( \frac{1}{2!}\right) ^{2}2\sum a_{n_{1}}^{\dagger
}b_{n_{1}}^{\dagger }a_{n_{2}}^{\dagger }b_{n_{2}}^{\dagger }\left\vert
v\right\rangle \left\langle v\right\vert b_{n_{2}}a_{n_{2}}b_{n_{1}}a_{n_{1}}
\end{equation*}%
\begin{equation}
=\frac{1}{2}\sum B_{n_{1}}^{\dagger }B_{n_{2}}^{\dagger }\left\vert
v\right\rangle \left\langle v\right\vert B_{n_{2}}B_{n_{1}}  \tag{6.6}
\end{equation}%
If we now rewrite free pairs in terms of Frenkel excitons according to Eq.
(2.18) and we use Eq. (2.20), the above closure relation leads to%
\begin{equation}
I\simeq \frac{1}{2!}\sum_{\mathbf{Q}_{1}\text{ }\mathbf{Q}_{2}}B_{\mathbf{Q}%
_{1}}^{\dagger }B_{\mathbf{Q}_{2}}^{\dagger }\left\vert v\right\rangle
\left\langle v\right\vert B_{\mathbf{Q}_{2}}B_{\mathbf{Q}_{1}}  \tag{6.7}
\end{equation}%
We see that this sum has the same prefactor $1/2!$ as the one for two
elementary bosons. This equality between prefactors can be understood by
noting that there is only one way to construct two Frenkel excitons out of
two pairs belonging to the $(\varepsilon _{e}+\varepsilon _{h}-\varepsilon
_{0})$\ subspace; so that these excitons behave as elementary bosons with
respect to this projection operator.

\subsection{Direct check}

It is possible to check the validity of the approximate closure relation for
two free pairs, given in Eq. (6.6), by inserting it in front of $%
B_{m_{1}}^{\dagger }B_{m_{2}}^{\dagger }\left\vert v\right\rangle $. Indeed
Eq. (5.1) for the scalar product of two-free-pair states readily gives%
\begin{equation}
\frac{1}{2}\sum B_{n_{1}}^{\dagger }B_{n_{2}}^{\dagger }\left\vert
v\right\rangle \left\langle v\right\vert
B_{n_{2}}B_{n_{1}}B_{m_{1}}^{\dagger }B_{m_{2}}^{\dagger }\left\vert
v\right\rangle =B_{m_{1}}^{\dagger }B_{m_{2}}^{\dagger }\left\vert
v\right\rangle   \tag{6.8}
\end{equation}%
for any $m_{1}\neq m_{2}$, as expected for a closure relation, the state $%
B_{m}^{\dagger 2}\left\vert v\right\rangle $ reducing to 0 since the same
site cannot accommodate two electrons.

The check of the approximate closure relation for two Frenkel excitons,
given in Eq. (6.7), is less trivial. Indeed, if we insert Eq. (6.7) in front
of $B_{\mathbf{P}_{1}}^{\dagger }B_{\mathbf{P}_{2}}^{\dagger }\left\vert
v\right\rangle $, we find, using Eq. (5.2) for the scalar product of
two-Frenkel exciton states,%
\begin{equation*}
\frac{1}{2}\sum_{\mathbf{Q}_{1}\mathbf{Q}_{2}}B_{\mathbf{Q}_{1}}^{\dagger
}B_{\mathbf{Q}_{2}}^{\dagger }\left\vert v\right\rangle \left\langle
v\right\vert B_{\mathbf{Q}_{2}}B_{\mathbf{Q}_{1}}B_{\mathbf{P}_{1}}^{\dagger
}B_{\mathbf{P}_{2}}^{\dagger }\left\vert v\right\rangle =B_{\mathbf{P}%
_{1}}^{\dagger }B_{\mathbf{P}_{2}}^{\dagger }\left\vert v\right\rangle 
\end{equation*}%
\begin{equation}
-\sum_{\mathbf{Q}_{1}\text{\textbf{\ }}\mathbf{Q}_{2}}\lambda \left( _{%
\mathbf{Q}_{1}\ \ \mathbf{P}_{1}}^{\mathbf{Q}_{2}\ \ \mathbf{P}_{2}}\right)
B_{\mathbf{Q}_{1}}^{\dagger }B_{\mathbf{Q}_{2}}^{\dagger }\left\vert
v\right\rangle   \tag{6.9}
\end{equation}%
In order to show that the above sum reduces to zero as necessary to prove
that Eq. (6.7) is indeed a closure relation, we use Eq. (3.11) for $\lambda
\left( _{\mathbf{Q}_{1}\ \ \mathbf{P}_{1}}^{\mathbf{Q}_{2}\ \ \mathbf{P}%
_{2}}\right) $. We find

\begin{equation}
\sum_{\mathbf{Q}_{1}\text{\textbf{\ }}\mathbf{Q}_{2}}\lambda \left( _{%
\mathbf{Q}_{1}\ \ \mathbf{P}_{1}}^{\mathbf{Q}_{2}\ \ \mathbf{P}_{2}}\right)
B_{\mathbf{Q}_{1}}^{\dagger }B_{\mathbf{Q}_{2}}^{\dagger }=\frac{1}{N_{s}}%
\sum_{\mathbf{Q}}B_{\mathbf{P}_{1}+\mathbf{Q}}^{\dagger }B_{\mathbf{P}_{2}-%
\mathbf{Q}}^{\dagger }\left\vert v\right\rangle   \tag{6.10}
\end{equation}%
We then rewrite exciton operators in terms of free pairs according to Eq.
(2.17) and we use Eq. (2.19). This leads to%
\begin{equation*}
\sum_{\mathbf{Q}}B_{\mathbf{P}_{1}+\mathbf{Q}}^{\dagger }B_{\mathbf{P}_{2}-%
\mathbf{Q}}^{\dagger }=\sum_{n_{1}n_{2}}e^{i(\mathbf{P}_{1}\mathbf{.R}%
_{n_{1}}+\mathbf{P}_{2}\mathbf{.R}_{n_{2}})}B_{n_{1}}^{\dagger
}B_{n_{2}}^{\dagger }\sum_{\mathbf{Q}}e^{i\mathbf{Q.(\mathbf{R}}_{n_{1}}-%
\mathbf{\mathbf{R}}_{n_{2}}\mathbf{)}}
\end{equation*}%
\begin{equation}
=N_{s}\sum_{n_{1}}e^{i\mathbf{(\mathbf{P}_{1}-\mathbf{P}_{2}).\mathbf{R}}%
_{n_{1}}}B_{n_{1}}^{\dagger 2}  \tag{6.11}
\end{equation}%
So that since, $B_{n}^{2}=0$, due to Eq. (2.9), we end with

\begin{equation}
\sum_{\mathbf{Q}_{1}\text{\textbf{\ }}\mathbf{Q}_{2}}\lambda \left( _{%
\mathbf{Q}_{1}\ \ \mathbf{P}_{1}}^{\mathbf{Q}_{2}\ \ \mathbf{P}_{2}}\right)
B_{\mathbf{Q}_{1}}^{\dagger }B_{\mathbf{Q}_{2}}^{\dagger }=0  \tag{6.12}
\end{equation}%
This result must be physically related to the fact that there is only one
way to write two Frenkel excitons out of two pairs: If we try to exchange
the carriers of two Frenkel excitons as in the sum (6.12), we get zero.

\subsection{Closure relation for $N$ Frenkel excitons}

A similar procedure performed for N pairs leads to an approximate closure
relation for $N$ Frenkel excitons which reads%
\begin{equation}
I\simeq \frac{1}{N!}\sum B_{\mathbf{Q}_{1}}^{\dagger }\ldots B_{\mathbf{Q}%
_{N}}^{\dagger }\left\vert v\right\rangle \left\langle v\right\vert B_{%
\mathbf{Q}_{N}}\ldots B_{\mathbf{Q}_{1}}  \tag{6.13}
\end{equation}%
with again the same $1/N!$\ prefactor as the one for $N$ elementary bosons.

\section{Creation potential for Frenkel excitons}

Frenkel excitons feel each other through exchanges as described in the
previous sections. This "interaction", generated by the Pauli exclusion
principle, is of course unusual in the sense that it gives rise to
dimensionless Pauli scatterings as shown in Section III. Frenkel excitons
also interact through the Coulomb potential which exists between their
fermionic components. As for Wannier excitons \cite{8}, there is no clean
way to write Coulomb interactions between two excitons as a potential due to
the composite nature of the particles. Here also, we are going to overcome
this difficulty by introducing a set of "creation potentials". From them, it
becomes formally possible to derive the interaction scatterings of two
Frenkel excitons, as we shown in the next section.

\subsection{Definition}

The creation potential $V_{\mathbf{Q}}^{\dagger }$ of the Frenkel exciton $%
\mathbf{Q}$ is defined through 
\begin{equation}
\left[ H_{X},B_{\mathbf{Q}}^{\dagger }\right] =E_{\mathbf{Q}}B_{\mathbf{Q}%
}^{\dagger }+V_{\mathbf{Q}}^{\dagger }  \tag{7.1}
\end{equation}%
where $H_{X}=H_{X}^{(0)}+V_{coul}$ is the interacting exciton Hamiltonian
defined in Eq. (2.24) while $H_{X}^{(0)}=H_{eh}+S_{X}$ with $H_{eh}$, $S_{X}$
and $V_{coul}$\ given by Eqs.\ (2.3, 22, 25). To get $V_{\mathbf{Q}%
}^{\dagger }$ we have to calculate the commutator of the Frenkel exciton
creation operator $B_{\mathbf{Q}}^{\dagger }$ with each of the three terms
of $H_{X}$. However, since $\left( H_{X}^{(0)}-E_{\mathbf{Q}}\right) B_{%
\mathbf{Q}}^{\dagger }\left\vert v\right\rangle =0$, the $E_{\mathbf{Q}}B_{%
\mathbf{Q}}^{\dagger }$ part in the RHS of Eq. (7.1) must come from $\left[
H_{X}^{(0)},\text{ }B_{\mathbf{Q}}^{\dagger }\right] $. This leads us to
split Eq. (7.1) as%
\begin{equation}
\left[ H_{X}^{(0)},B_{\mathbf{Q}}^{\dagger }\right] =E_{\mathbf{Q}}B_{%
\mathbf{Q}}^{\dagger }+S_{\mathbf{Q}}^{\dagger }  \tag{7.2}
\end{equation}%
\begin{equation}
\left[ V_{coul},B_{\mathbf{Q}}^{\dagger }\right] =W_{\mathbf{Q}}^{\dagger } 
\tag{7.3}
\end{equation}%
The Frenkel exciton creation potential is thus going to appear as%
\begin{equation}
V_{\mathbf{Q}}^{\dagger }=S_{\mathbf{Q}}^{\dagger }+W_{\mathbf{Q}}^{\dagger }
\tag{7.4}
\end{equation}%
these two parts being conceptually quite different: $W_{\mathbf{Q}}^{\dagger
}$ associated to $V_{coul}$ comes from direct Coulomb processes between
sites. It has similarity with the Wannier exciton creation potential. In
contrast, $S_{\mathbf{Q}}^{\dagger }$, which originates from $S_{X}$,
essentially comes from indirect Coulomb processes; these processes, crucial
for Frenkel excitons as they are responsible for the excitation transfer,
are usually neglected for Wannier excitons. They precisely correspond to the
so-called "electron-hole exchange".

\subsection{Free exciton creation potential}

The free exciton Hamiltonian $H_{X}^{(0)}$ contains an electron - hole
kinetic energy term $H_{eh}$ defined in Eq. (2.3). Its commutator with $B_{%
\mathbf{Q}}^{\dagger }$ readily gives%
\begin{equation*}
\left[ H_{eh},B_{\mathbf{Q}}^{\dagger }\right] =\left[ \varepsilon _{e}\sum
a_{n}^{\dagger }a_{n}+\varepsilon _{h}\sum b_{n}^{\dagger }b_{n},\text{ }%
\sum \frac{e^{i\mathbf{Q}.\mathbf{R}_{m}}}{\sqrt{N_{s}}}a_{m}^{\dagger
}b_{m}^{\dagger }\right] 
\end{equation*}%
\begin{equation}
=(\varepsilon _{e}+\varepsilon _{h})B_{\mathbf{Q}}^{\dagger }  \tag{7.5}
\end{equation}

$H_{X}^{(0)}$ also contains Coulomb contributions through $S_{X}$. Although $%
S_{X}$ reads as a \textit{one}-body diagonal potential in the Frenkel
exciton subspace, it can produce interactions between \textit{two} excitons
through the carrier exchanges the exciton composite nature allows - an idea
not trivial at first.

By using Eq. (3.6), it is straightforward to see that $\left[ S_{X},B_{%
\mathbf{Q}}^{\dagger }\right] $ reads in terms of the deviation-from-boson
operator for Frenkel exciton as%
\begin{equation*}
\left[ S_{X},B_{\mathbf{Q}}^{\dagger }\right] =\left[ \sum_{\mathbf{Q}%
^{\prime }}\zeta _{\mathbf{Q}^{\prime }}B_{\mathbf{Q}^{\prime }}^{\dagger
}B_{\mathbf{Q}^{\prime }},\text{ }B_{\mathbf{Q}}^{\dagger }\right] =\sum_{%
\mathbf{Q}^{\prime }}\zeta _{\mathbf{Q}^{\prime }}B_{\mathbf{Q}^{\prime
}}^{\dagger }\left( \delta _{\mathbf{Q}^{\prime }\text{ }\mathbf{Q}}-D_{%
\mathbf{Q}^{\prime }\mathbf{Q}}\right) 
\end{equation*}%
\begin{equation}
=\zeta _{\mathbf{Q}}B_{\mathbf{Q}}^{\dagger }-\sum_{\mathbf{Q}^{\prime
}}\zeta _{\mathbf{Q}^{\prime }}B_{\mathbf{Q}^{\prime }}^{\dagger }D_{\mathbf{%
Q}^{\prime }\mathbf{Q}}  \tag{7.6}
\end{equation}%
When combined with Eq. (7.5), we readily get Eq. (7.2), with $S_{\mathbf{Q}%
}^{\dagger }$ given by%
\begin{equation}
S_{\mathbf{Q}}^{\dagger }=-\sum_{\mathbf{Q}^{\prime }}\zeta _{\mathbf{Q}%
^{\prime }}B_{\mathbf{Q}^{\prime }}^{\dagger }D_{\mathbf{Q}^{\prime }\mathbf{%
Q}}=-\sum_{\mathbf{Q}^{\prime }}B_{\mathbf{Q}^{\prime }}^{\dagger }\zeta _{%
\mathbf{Q}^{\prime }}\left( \Delta _{\mathbf{Q}-\mathbf{Q}^{\prime
}}^{(e)}+\Delta _{\mathbf{Q}-\mathbf{Q}^{\prime }}^{(h)}\right)   \tag{7.7}
\end{equation}

$\zeta _{\mathbf{Q}}$ appearing in $S_{\mathbf{Q}}^{\dagger }$ is defined in
Eq. (2.23). It has a constant contribution $(-\varepsilon _{0})$ which comes
from the neutrality potential $V_{intra}$ induced by intrasite Coulomb
interactions. It also has a contribution $\mathcal{V}_{\mathbf{Q}}$ which
comes from electron-hole exchange interactions between sites which allow to
delocalize the excitation. As shown below, the interaction scatterings
generated by these two contributions have rather different properties; this
is why it is physically relevant to split $S_{\mathbf{Q}}^{\dagger }$ as%
\begin{equation}
S_{\mathbf{Q}}^{\dagger }=N_{\mathbf{Q}}^{\dagger }+T_{\mathbf{Q}}^{\dagger }
\tag{7.8}
\end{equation}%
where $N_{\mathbf{Q}}^{\dagger }$ and $T_{\mathbf{Q}}^{\dagger }$ read as $%
S_{\mathbf{Q}}^{\dagger }$ in Eq. (7.7), with $\zeta _{\mathbf{Q}^{\prime }}$
replaced by $(-\varepsilon _{0})$ and $\mathcal{V}_{\mathbf{Q}^{\prime }}$
respectively.

\subsection{Interacting exciton creation potential}

We now turn to the creation potential induced by direct processes through $%
V_{coul}$. This interaction contains three parts. The electron-electron
contribution gives%
\begin{equation}
\left[ V_{ee},\text{ }B_{\mathbf{Q}}^{\dagger }\right] =\left[ \frac{1}{2}%
\sum_{n_{1}\neq n_{2}}V_{\mathbf{R}_{n_{1}}-\mathbf{R}_{n_{2}}}\left( _{1\
1}^{1\ 1}\right) a_{n_{1}}^{\dagger }a_{n_{2}}^{\dagger }a_{n_{2}}a_{n_{1}},%
\text{ }\sum_{m}\frac{e^{i\mathbf{Q}.\mathbf{R}_{m}}}{\sqrt{N_{s}}}%
a_{m}^{\dagger }b_{m}^{\dagger }\right]  \tag{7.9}
\end{equation}%
By using $\left[ a_{n_{1}}^{\dagger }a_{n_{2}}^{\dagger }a_{n_{2}}a_{n_{1}},%
\text{ }a_{m}^{\dagger }b_{m}^{\dagger }\right] =a_{m}^{\dagger
}b_{m}^{\dagger }\left( \delta _{n_{1}m}a_{n_{2}}^{\dagger }a_{n_{2}}+\delta
_{n_{2}m}a_{n_{1}}^{\dagger }a_{n_{1}}\right) $ and by writing $%
a_{m}^{\dagger }b_{m}^{\dagger }$ in terms of excitons through Eq. (3.10),
we find, since $V_{\mathbf{R}}\left( _{1\ 1}^{1\ 1}\right) =V_{-\mathbf{R}%
}\left( _{1\ 1}^{1\ 1}\right) $ 
\begin{equation}
\left[ V_{ee},\text{ }B_{\mathbf{Q}}^{\dagger }\right] =\sum_{\mathbf{Q}%
^{\prime }}B_{\mathbf{Q}^{\prime }}^{\dagger }\sum_{n}a_{n}^{\dagger
}a_{n}e^{i\mathbf{(\mathbf{Q-Q}^{\prime })}.\mathbf{\mathbf{R}}%
_{n}}\sum_{m\neq n}\frac{e^{i\mathbf{(\mathbf{Q-Q}^{\prime })}.\mathbf{(%
\mathbf{R}}_{m}\mathbf{-\mathbf{R}}_{n}\mathbf{)}}}{N_{s}}V_{\mathbf{R}_{n}-%
\mathbf{R}_{m}}\left( _{1\ 1}^{1\ 1}\right)  \tag{7.10}
\end{equation}%
The last sum does not depend on $n$ due to the translational invariance of
the system; so that the above equation can be rewritten as%
\begin{equation}
\left[ V_{ee},\text{ }B_{\mathbf{Q}}^{\dagger }\right] =\sum_{\mathbf{Q}%
^{\prime }}B_{\mathbf{Q}^{\prime }}^{\dagger }\mathcal{W}_{\mathbf{\mathbf{%
Q-Q}^{\prime }}}^{(ee)}\Delta _{\mathbf{Q}-\mathbf{Q}^{\prime }}^{(e)} 
\tag{7.11}
\end{equation}%
where $\mathcal{W}_{\mathbf{\mathbf{Q}}}^{(ee)}$ is very similar to the $%
\mathbf{\mathbf{Q}}$ component of the exciton energy $\mathcal{V}_{\mathbf{%
\mathbf{Q}}}$ given in Eq. (2.15), except that it reads in terms of direct
Coulomb interactions between two excited atomic levels $\nu =1$. Its precise
value is given by%
\begin{equation}
\mathcal{W}_{\mathbf{\mathbf{Q}}}^{(ee)}=\sum_{\mathbf{R\neq }0}e^{-i\mathbf{%
\mathbf{Q}}.\mathbf{R}}V_{\mathbf{R}}\left( _{1\ 1}^{1\ 1}\right)  \tag{7.12}
\end{equation}

In the same way, the hole-hole contribution of $V_{coul}$ leads to%
\begin{equation}
\left[ V_{hh},\text{ }B_{\mathbf{Q}}^{\dagger }\right] =\sum_{\mathbf{Q}%
^{\prime }}B_{\mathbf{Q}^{\prime }}^{\dagger }\mathcal{W}_{\mathbf{\mathbf{%
Q-Q}^{\prime }}}^{(hh)}\Delta _{\mathbf{Q}-\mathbf{Q}^{\prime }}^{(h)} 
\tag{7.13}
\end{equation}%
where $\mathcal{W}_{\mathbf{\mathbf{Q}}}^{(hh)}$ reads as $\mathcal{W}_{%
\mathbf{\mathbf{Q}}}^{(ee)}$\ with $V_{\mathbf{R}}\left( _{1\ 1}^{1\
1}\right) $ replaced by $V_{\mathbf{R}}\left( _{0\ 0}^{0\ 0}\right) $.

If we now turn to electron-hole direct interaction, its commutator with $B_{%
\mathbf{Q}}^{\dagger }$ gives%
\begin{equation}
\left[ V_{eh}^{(dir)},\text{ }B_{\mathbf{Q}}^{\dagger }\right] =\left[
-\sum_{n\neq n^{\prime }}V_{\mathbf{R}_{n}-\mathbf{R}_{n^{\prime }}}\left(
_{0\ 0}^{1\ 1}\right) b_{n}^{\dagger }a_{n^{\prime }}^{\dagger }a_{n^{\prime
}}b_{n},\text{ }\sum_{m}\frac{e^{i\mathbf{Q}.\mathbf{R}_{m}}}{\sqrt{N_{s}}}%
a_{m}^{\dagger }b_{m}^{\dagger }\right]  \tag{7.14}
\end{equation}%
To calculate it, we use $\left[ b_{n}^{\dagger }a_{n^{\prime }}^{\dagger
}a_{n^{\prime }}b_{n},\text{ }a_{m}^{\dagger }b_{m}^{\dagger }\right]
=a_{m}^{\dagger }b_{m}^{\dagger }(\delta _{nm}\delta _{n^{\prime }m}+\delta
_{nm}a_{n^{\prime }}^{\dagger }a_{n^{\prime }}+\delta _{n^{\prime
}m}b_{n}^{\dagger }b_{n}).$ The first term of this commutator gives 0 for $%
n\neq n^{\prime }$. In the two other terms, we rewrite $a_{m}^{\dagger
}b_{m}^{\dagger }$ in terms of excitons according to Eq. (2.18) and we make
appearing $e^{i\mathbf{(Q-Q}^{\prime }\mathbf{)}.\mathbf{(\mathbf{R}}_{m}%
\mathbf{-\mathbf{R}}_{n}\mathbf{)}}$ in the remaining sum to possibly use
the translational invariance of the system. This ultimately leads to 
\begin{equation}
\left[ V_{eh}^{(dir)},\text{ }B_{\mathbf{Q}}^{\dagger }\right] =-\sum_{%
\mathbf{Q}^{\prime }}B_{\mathbf{Q}^{\prime }}^{\dagger }\sum_{n}\left( 
\mathcal{W}_{\mathbf{\mathbf{Q-Q}^{\prime }}}^{(eh)}\Delta _{\mathbf{Q}-%
\mathbf{Q}^{\prime }}^{(e)}+\mathcal{W}_{\mathbf{\mathbf{Q-Q}^{\prime }}%
}^{(he)}\Delta _{\mathbf{Q}-\mathbf{Q}^{\prime }}^{(h)}\right)  \tag{7.15}
\end{equation}%
where $\mathcal{W}_{\mathbf{\mathbf{Q}}}^{(eh)}$ and $\mathcal{W}_{\mathbf{%
\mathbf{Q}}}^{(he)}$ read as $\mathcal{W}_{\mathbf{\mathbf{Q}}}^{(ee)}$
given in Eq. (7.12) with $V_{\mathbf{R}}\left( _{1\ 1}^{1\ 1}\right) $
replaced by $V_{\mathbf{R}}\left( _{0\ 0}^{1\ 1}\right) $ and $V_{\mathbf{R}%
}\left( _{1\ 1}^{0\ 0}\right) $ respectively.

By collecting these three contributions, we find that $W_{\mathbf{Q}%
}^{\dagger }=\left[ V_{coul,}B_{\mathbf{Q}}^{\dagger }\right] $ reads as 
\begin{equation}
W_{\mathbf{Q}}^{\dagger }=\sum_{\mathbf{Q}^{\prime }}B_{\mathbf{Q}^{\prime
}}^{\dagger }\left( \mathcal{W}_{\mathbf{\mathbf{Q-Q}^{\prime }}%
}^{(e)}\Delta _{\mathbf{Q}-\mathbf{Q}^{\prime }}^{(e)}+\mathcal{W}_{\mathbf{%
\mathbf{Q-Q}^{\prime }}}^{(h)}\Delta _{\mathbf{Q}-\mathbf{Q}^{\prime
}}^{(h)}\right)  \tag{7.16}
\end{equation}%
where the prefactors are such that $\mathcal{W}_{\mathbf{\mathbf{Q}}}^{(e)}=%
\mathcal{W}_{\mathbf{\mathbf{Q}}}^{(ee)}-\mathcal{W}_{\mathbf{\mathbf{Q}}%
}^{(eh)}$ while $\mathcal{W}_{\mathbf{\mathbf{Q}}}^{(h)}=\mathcal{W}_{%
\mathbf{\mathbf{Q}}}^{(hh)}-\mathcal{W}_{\mathbf{\mathbf{Q}}}^{(he)}$; so
that 
\begin{equation}
\mathcal{W}_{\mathbf{\mathbf{Q}}}^{(e)}=\sum_{\mathbf{R\neq 0}}e^{-i\mathbf{%
\mathbf{Q}}.\mathbf{R}}\left[ V_{\mathbf{R}}\left( _{1\ 1}^{1\ 1}\right) -V_{%
\mathbf{R}}\left( _{0\ 0}^{1\ 1}\right) \right]  \tag{7.17}
\end{equation}%
\begin{equation}
\mathcal{W}_{\mathbf{\mathbf{Q}}}^{(h)}=\sum_{\mathbf{R\neq 0}}e^{-i\mathbf{%
\mathbf{Q}}.\mathbf{R}}\left[ V_{\mathbf{R}}\left( _{0\ 0}^{0\ 0}\right) -V_{%
\mathbf{R}}\left( _{1\ 1}^{0\ 0}\right) \right]  \tag{7.18}
\end{equation}

\subsection{Discussion}

It is of importance to note that, while the structure of the two parts $S_{%
\mathbf{Q}}^{\dagger }$ and $W_{\mathbf{Q}}^{\dagger }$ of the Frenkel
exciton creation potential $V_{\mathbf{Q}}^{\dagger }$ are rather similar,
the prefactors of $B_{\mathbf{Q}^{\prime }}^{\dagger }$\ in $W_{\mathbf{Q}%
}^{\dagger }$\ which come from \textit{direct} Coulomb processes between
electrons and holes of different sites, depend on the momentum transfer $%
\mathbf{\mathbf{Q-Q}^{\prime }}$\ only, while this is not true for the
prefactor of $B_{\mathbf{Q}^{\prime }}^{\dagger }$ in $S_{\mathbf{Q}%
}^{\dagger }$ which comes from "electron-hole exchange", i.e., \textit{%
indirect} Coulomb processes between atomic levels. This result, which may
appear as very strange at first, is linked to the fact that these indirect
Coulomb processes do not lead to a momentum transfer: Indeed, the potential $%
S_{X}$ they produce is a one-body \textit{diagonal} operator in the exciton
subspace, as seen from Eq. (2.22).

Let us recall that the Coulomb exchange processes producing $S_{\mathbf{Q}%
}^{\dagger }$ in which one electron is excited while the other returns to
its ground state, are usually neglected in the case of Wannier excitons.
Being much smaller than the direct scatterings due to the orthogonality of
the ground and excited state wave functions, they just produce the small
energy splitting which exists between dark and bright excitons, when the
spin degrees of freedom of these Wannier excitons are included. In the case
of Frenkel excitons, the prefactors in $S_{\mathbf{Q}}^{\dagger }$ are also
small compared to the ones coming from direct processes since through $V_{%
\mathbf{R}}\left( _{1\ 0}^{0\ 1}\right) $, they also contain overlaps of
different atomic states $\nu =0$ and $\nu =1$ - which are orthogonal - while
the prefactors of $W_{\mathbf{Q}}^{\dagger }$ contain $V_{\mathbf{R}}\left(
_{\nu ^{\prime }\ \nu ^{\prime }}^{\nu \ \ \ \nu }\right) $, i.e., overlaps
of identical atomic states, $\nu $ or $\nu ^{\prime }$. However, these
indirect Coulomb processes are crucial for Frenkel excitons since, in the
tight binding limit, they are the only processes allowing to delocalize the
excitation, as necessary to produce the exciton.

\section{Interaction scatterings}

\subsection{Definition}

The interaction scatterings of the two Frenkel excitons ($\mathbf{Q}_{1}$, $%
\mathbf{Q}_{2}$) are defined through the creation potential $V_{\mathbf{Q}%
}^{\dagger }$ as 
\begin{equation}
\left[ V_{\mathbf{Q}_{1}}^{\dagger },B_{\mathbf{Q}_{2}}^{\dagger }\right]
=\sum\limits_{\mathbf{Q}_{1}^{\prime }\text{ }\mathbf{Q}_{2}^{\prime }}\xi
\left( _{\mathbf{Q}_{1}^{\prime }\ \mathbf{Q}_{1}}^{\mathbf{Q}_{2}^{\prime
}\ \mathbf{Q}_{2}}\right) B_{\mathbf{Q}_{1}^{\prime }}^{\dagger }B_{\mathbf{Q%
}_{2}^{\prime }}^{\dagger }  \tag{8.1}
\end{equation}

Before going further, let us recall that this equation, when used for
Wannier excitons, does not unambiguously define \cite{14} the interaction
scattering $\xi \left( _{m\ i}^{n\ j}\right) $ between ($i$, $j$) and ($m$, $%
n$) states. Indeed, due to the two possible ways to write two excitons
leading to Eq. (4.8), it is always possible to replace $\xi \left( _{m\
i}^{n\ j}\right) $ in Eq. (8.1) by minus the "in" exchange interaction
scattering defined as%
\begin{equation}
\xi ^{in}\left( _{m\ i}^{n\ j}\right) =\sum\limits_{p\text{ }q}\lambda
\left( _{m\ p}^{n\ q}\right) \xi \left( _{p\ i}^{q\ j}\right)  \tag{8.2}
\end{equation}%
or even by $\left[ a\xi \left( _{m\ i}^{n\ j}\right) -b\xi ^{in}\left( _{m\
i}^{n\ j}\right) \right] $ with $a+b=1$.

In contrast, due to the fact that there is only one way to construct two
Frenkel excitons out of two electron-hole pairs, an equation like Eq. (4.8)
does not exist for Frenkel excitons. This leads us to think that the value
of the interaction scattering $\xi \left( _{\mathbf{Q}_{1}^{\prime }\ 
\mathbf{Q}_{1}}^{\mathbf{Q}_{2}^{\prime }\ \mathbf{Q}_{2}}\right) $ obtained
through Eq. (8.1) must be unique. As a direct consequence, the "in" exchange
interaction scattering for Frenkel excitons must be equal to zero. A way to
show it, is to insert the closure relation (6.7) for two Frenkel exciton
states in front of RHS of Eq. (8.1) acting on vacuum. We find%
\begin{equation}
\sum\limits_{\mathbf{Q}_{1}^{\prime }\text{ }\mathbf{Q}_{2}^{\prime }}\xi
\left( _{\mathbf{Q}_{1}^{\prime }\ \mathbf{Q}_{1}}^{\mathbf{Q}_{2}^{\prime
}\ \mathbf{Q}_{2}}\right) B_{\mathbf{Q}_{1}^{\prime }}^{\dagger }B_{\mathbf{Q%
}_{2}^{\prime }}^{\dagger }\left\vert v\right\rangle =\sum\limits_{\mathbf{P}%
_{1}\text{ }\mathbf{P}_{2}}\widehat{\xi }\left( _{\mathbf{P}_{1}\ \mathbf{Q}%
_{1}}^{\mathbf{P}_{2}\ \mathbf{Q}_{2}}\right) B_{\mathbf{P}_{1}}^{\dagger
}B_{\mathbf{P}_{2}}^{\dagger }\left\vert v\right\rangle  \tag{8.3}
\end{equation}%
where the scattering appearing in the sum is given by%
\begin{equation}
\widehat{\xi }\left( _{\mathbf{P}_{1}\ \mathbf{Q}_{1}}^{\mathbf{P}_{2}\ 
\mathbf{Q}_{2}}\right) =\frac{1}{2}\sum\limits_{\mathbf{Q}_{1}^{\prime }%
\text{ }\mathbf{Q}_{2}^{\prime }}\xi \left( _{\mathbf{Q}_{1}^{\prime }\ 
\mathbf{Q}_{1}}^{\mathbf{Q}_{2}^{\prime }\ \mathbf{Q}_{2}}\right)
\left\langle v\right\vert B_{\mathbf{P}_{2}}B_{\mathbf{P}_{1}}B_{\mathbf{Q}%
_{1}}^{\dagger }B_{\mathbf{Q}_{2}}^{\dagger }\left\vert v\right\rangle 
\tag{8.4}
\end{equation}%
If we now use Eq. (5.2) for the scalar product of two Frenkel exciton states
and we symmetrize the interaction scattering, $\xi \left( _{\mathbf{Q}%
_{1}^{\prime }\ \mathbf{Q}_{1}}^{\mathbf{Q}_{2}^{\prime }\ \mathbf{Q}%
_{2}}\right) =\xi \left( _{\mathbf{Q}_{2}^{\prime }\ \mathbf{Q}_{1}}^{%
\mathbf{Q}_{1}^{\prime }\ \mathbf{Q}_{2}}\right) $, as always possible since 
$B_{\mathbf{Q}_{1}^{\prime }}^{\dagger }B_{\mathbf{Q}_{2}^{\prime
}}^{\dagger }=B_{\mathbf{Q}_{2}^{\prime }}^{\dagger }B_{\mathbf{Q}%
_{1}^{\prime }}^{\dagger }$, we find that $\widehat{\xi }\left( _{\mathbf{P}%
_{1}\ \mathbf{Q}_{1}}^{\mathbf{P}_{2}\ \mathbf{Q}_{2}}\right) $ is just%
\begin{equation}
\widehat{\xi }\left( _{\mathbf{P}_{1}\ \mathbf{Q}_{1}}^{\mathbf{P}_{2}\ 
\mathbf{Q}_{2}}\right) =\xi \left( _{\mathbf{P}_{1}\ \mathbf{Q}_{1}}^{%
\mathbf{P}_{2}\ \mathbf{Q}_{2}}\right) -\xi ^{in}\left( _{\mathbf{P}_{1}\ 
\mathbf{Q}_{1}}^{\mathbf{P}_{2}\ \mathbf{Q}_{2}}\right)  \tag{8.5}
\end{equation}%
where $\xi ^{in}\left( _{\mathbf{P}_{1}\ \mathbf{Q}_{1}}^{\mathbf{P}_{2}\ 
\mathbf{Q}_{2}}\right) $ is the "in" exchange interaction scattering of two
Frenkel excitons, defined in the same way as for Wannier excitons, namely%
\begin{equation}
\xi ^{in}\left( _{\mathbf{P}_{1}\ \mathbf{Q}_{1}}^{\mathbf{P}_{2}\ \mathbf{Q}%
_{2}}\right) =\sum\limits_{\mathbf{Q}_{1}^{\prime }\text{ }\mathbf{Q}%
_{2}^{\prime }}\lambda \left( _{\mathbf{P}_{1}\ \mathbf{Q}_{1}^{\prime }}^{%
\mathbf{P}_{2}\ \mathbf{Q}_{2}^{\prime }}\right) \xi \left( _{\mathbf{Q}%
_{1}^{\prime }\ \mathbf{Q}_{1}}^{\mathbf{Q}_{2}^{\prime }\ \mathbf{Q}%
_{2}}\right)  \tag{8.6}
\end{equation}%
By inserting $\widehat{\xi }$ given in Eq. (8.5) into Eq. (8.3), we are led
to conclude that $\xi ^{in}\left( _{\mathbf{P}_{1}\ \mathbf{Q}_{1}}^{\mathbf{%
P}_{2}\ \mathbf{Q}_{2}}\right) =0$, or if $\xi ^{in}$ has a non zero part,
its contribution to the sum (8.3) must give zero.

Let us now see how this conclusion, based on very general properties of
Frenkel excitons, can be recovered through hard algebra.

\subsection{Calculation of the interaction scattering}

To calculate the interaction scattering $\xi \left( _{\mathbf{Q}_{1}^{\prime
}\ \mathbf{Q}_{1}}^{\mathbf{Q}_{2}^{\prime }\ \mathbf{Q}_{2}}\right) $\
defined in Eq. (8.1), with the creation potential $V_{\mathbf{Q}}^{\dagger }$
given in Eq. (7.4), we first note that%
\begin{equation*}
\left[ \Delta _{\mathbf{P}}^{(e)},B_{\mathbf{Q}}^{\dagger }\right] =\left[ 
\frac{1}{N_{s}}\sum_{n}e^{i\mathbf{P}.\mathbf{R}_{n}}a_{n}^{\dagger }a_{n},%
\frac{1}{\sqrt{N_{s}}}\sum_{m}e^{i\mathbf{Q}.\mathbf{R}_{m}}a_{m}^{\dagger
}b_{m}^{\dagger }\right] 
\end{equation*}%
\begin{equation}
=\frac{1}{N_{s}\sqrt{N_{s}}}\sum_{m}e^{i\mathbf{(P+Q)}.\mathbf{R}%
_{m}}a_{m}^{\dagger }b_{m}^{\dagger }  \tag{8.7}
\end{equation}%
So that, by writing free electron-hole operators in terms of excitons
according to Eq. (2.18), and by using Eq. (2.20), we end with%
\begin{equation}
\left[ \Delta _{\mathbf{P}}^{(e)},B_{\mathbf{Q}}^{\dagger }\right] =\frac{1}{%
N_{s}}B_{\mathbf{P}+\mathbf{Q}}^{\dagger }=\left[ \Delta _{\mathbf{P}%
}^{(h)},B_{\mathbf{Q}}^{\dagger }\right]   \tag{8.8}
\end{equation}%
This readily shows that 
\begin{equation}
\left[ S_{\mathbf{Q}_{1}}^{\dagger },B_{\mathbf{Q}_{2}}^{\dagger }\right] =-%
\frac{2}{N_{s}}\sum_{\mathbf{Q}_{1}^{\prime }}\zeta _{\mathbf{Q}_{1}^{\prime
}}B_{\mathbf{Q}_{1}^{\prime }}^{\dagger }B_{\mathbf{Q}_{1}+\mathbf{Q}_{2}-%
\mathbf{Q}_{1}^{\prime }}^{\dagger }  \tag{8.9}
\end{equation}%
\begin{equation}
\left[ W_{\mathbf{Q}_{1}}^{\dagger },B_{\mathbf{Q}_{2}}^{\dagger }\right] =%
\frac{1}{N_{s}}\sum_{\mathbf{Q}_{1}^{\prime }}\mathcal{W}_{\mathbf{Q}_{1}-%
\mathbf{Q}_{1}^{\prime }}B_{\mathbf{Q}_{1}^{\prime }}^{\dagger }B_{\mathbf{Q}%
_{1}+\mathbf{Q}_{2}-\mathbf{Q}_{1}^{\prime }}^{\dagger }  \tag{8.10}
\end{equation}%
where $\zeta _{\mathbf{Q}}$ is defined in Eq. (2.23) while $\mathcal{W}_{%
\mathbf{Q}}$ is equal to $\mathcal{W}_{\mathbf{Q}}^{(e)}+\mathcal{W}_{%
\mathbf{Q}}^{(h)}$; so that%
\begin{equation*}
\mathcal{W}_{\mathbf{Q}}=\sum_{\mathbf{R\neq 0}}e^{-i\mathbf{\mathbf{Q}}.%
\mathbf{R}}\left[ V_{\mathbf{R}}\left( _{1\ 1}^{1\ 1}\right) +V_{\mathbf{R}%
}\left( _{0\ 0}^{0\ 0}\right) -V_{\mathbf{R}}\left( _{0\ 0}^{1\ 1}\right)
-V_{\mathbf{R}}\left( _{1\ 1}^{0\ 0}\right) \right] 
\end{equation*}
\begin{equation}
=\mathcal{W}_{\mathbf{Q}}^{\ast }=\mathcal{W}_{-\mathbf{Q}}  \tag{8.11}
\end{equation}%
It contains all possible direct interactions between electrons and holes in
different sites. This quantity is real, since $V_{\mathbf{R}}\left( _{\nu \
\nu }^{\nu \ \nu }\right) =V_{-\mathbf{R}}\left( _{\nu \ \nu }^{\nu \ \nu
}\right) $ while $V_{\mathbf{R}}^{\ast }\left( _{0\ 0}^{1\ 1}\right) =V_{-%
\mathbf{R}}\left( _{1\ 1}^{0\ 0}\right) $. For the same reason, it is an
even function of $\mathbf{Q}$.

From Eqs. (7.4) and (7.8), we are lead to split the interaction scattering
defined in Eq. (8.1) into three conceptually different terms%
\begin{equation}
\xi \left( _{\mathbf{Q}_{1}^{\prime }\ \mathbf{Q}_{1}}^{\mathbf{Q}%
_{2}^{\prime }\ \mathbf{Q}_{2}}\right) =\xi _{coul}\left( _{\mathbf{Q}%
_{1}^{\prime }\ \mathbf{Q}_{1}}^{\mathbf{Q}_{2}^{\prime }\ \mathbf{Q}%
_{2}}\right) -\xi _{trans}\left( _{\mathbf{Q}_{1}^{\prime }\ \mathbf{Q}%
_{1}}^{\mathbf{Q}_{2}^{\prime }\ \mathbf{Q}_{2}}\right) +\xi _{neut}\left( _{%
\mathbf{Q}_{1}^{\prime }\ \mathbf{Q}_{1}}^{\mathbf{Q}_{2}^{\prime }\ \mathbf{%
Q}_{2}}\right)  \tag{8.12}
\end{equation}%
The first contribution which reads%
\begin{equation}
\xi _{coul}\left( _{\mathbf{Q}_{1}^{\prime }\ \mathbf{Q}_{1}}^{\mathbf{Q}%
_{2}^{\prime }\ \mathbf{Q}_{2}}\right) =\frac{1}{N_{s}}\mathcal{W}_{\mathbf{Q%
}_{1}-\mathbf{Q}_{1}^{\prime }}\delta _{\mathbf{Q}_{1}^{\prime }+\mathbf{Q}%
_{2}^{\prime }\text{ }\mathbf{Q}_{1}+\mathbf{Q}_{2}}  \tag{8.13}
\end{equation}%
only depends on the momentum transfer $\mathbf{Q}_{1}-\mathbf{Q}_{1}^{\prime
}$ produced by the scattering. Through $\mathcal{W}_{\mathbf{Q}}$, this
contribution comes from all \textit{direct} Coulomb processes between the
excited atomic level $\nu =1$\ and the ground state level $\nu =0$ of the
different sites. This part of the interaction scattering which is
represented by the diagram of Fig. 5(a), is very similar to the direct
Coulomb scattering of two Wannier excitons.

The two other contributions to the interaction scattering are conceptually
new. By symmetrizing Eq. (8.9), it is possible to write them as%
\begin{equation}
\xi _{transf}\left( _{\mathbf{Q}_{1}^{\prime }\ \ \mathbf{Q}_{1}}^{\mathbf{Q}%
_{2}^{\prime }\ \ \mathbf{Q}_{2}}\right) =\frac{1}{N_{s}}(\mathcal{V}_{%
\mathbf{Q}_{1}^{\prime }}+\mathcal{V}_{\mathbf{Q}_{2}^{\prime }})\delta _{%
\mathbf{Q}_{1}^{\prime }+\mathbf{Q}_{2}^{\prime },\mathbf{Q}_{1}+\mathbf{Q}%
_{2}}  \tag{8.14}
\end{equation}%
\begin{equation}
\xi _{neut}\left( _{\mathbf{Q}_{1}^{\prime }\ \ \mathbf{Q}_{1}}^{\mathbf{Q}%
_{2}^{\prime }\ \ \mathbf{Q}_{2}}\right) =\frac{2\varepsilon _{0}}{N_{s}}%
\delta _{\mathbf{Q}_{1}^{\prime }+\mathbf{Q}_{2}^{\prime },\mathbf{Q}_{1}+%
\mathbf{Q}_{2}}  \tag{8.15}
\end{equation}%
These scatterings may appear as very strange at first since they do not
depend on the momentum transfer but on the momenta of the two "out" excitons 
$\mathbf{Q}_{1}^{\prime }$ and $\mathbf{Q}_{2}^{\prime }$ for $\xi _{transf}$%
, while $\xi _{neut}$\ is completely independent of the momenta of the
Frenkel excitons involved in the scattering - which is even stranger.

Actually $\xi _{neut}$ does not play a role in the scattering of two Frenkel
excitons. Indeed, when inserted in Eq. (8.1), its contribution gives zero%
\begin{equation}
\sum \xi _{neut}\left( _{\mathbf{Q}_{1}^{\prime }\ \ \mathbf{Q}_{1}}^{%
\mathbf{Q}_{2}^{\prime }\ \ \mathbf{Q}_{2}}\right) B_{\mathbf{Q}_{1}^{\prime
}}^{\dagger }B_{\mathbf{Q}_{2}^{\prime }}^{\dagger }=0  \tag{8.16}
\end{equation}%
This can be readily shown by noting that%
\begin{equation}
\xi _{neut}\left( _{\mathbf{Q}_{1}^{\prime }\ \ \mathbf{Q}_{1}}^{\mathbf{Q}%
_{2}^{\prime }\ \ \mathbf{Q}_{2}}\right) =2\varepsilon _{0}\lambda \left( _{%
\mathbf{Q}_{1}^{\prime }\ \ \mathbf{Q}_{1}}^{\mathbf{Q}_{2}^{\prime }\ \ 
\mathbf{Q}_{2}}\right)  \tag{8.17}
\end{equation}%
and by using Eq. (6.12). Consequently $\xi $ in Eq. (8.1) can as well be
replaced by%
\begin{equation}
\widetilde{\xi }\left( _{\mathbf{Q}_{1}^{\prime }\ \ \mathbf{Q}_{1}}^{%
\mathbf{Q}_{2}^{\prime }\ \ \mathbf{Q}_{2}}\right) =\xi _{coul}\left( _{%
\mathbf{Q}_{1}^{\prime }\ \ \mathbf{Q}_{1}}^{\mathbf{Q}_{2}^{\prime }\ \ 
\mathbf{Q}_{2}}\right) -\xi _{trans}\left( _{\mathbf{Q}_{1}^{\prime }\ \ 
\mathbf{Q}_{1}}^{\mathbf{Q}_{2}^{\prime }\ \ \mathbf{Q}_{2}}\right) 
\tag{8.18}
\end{equation}

If we now turn to $\xi _{transf}$, given in Eq. (8.14), we can note that,
due to Eq. (3.11), it also reads%
\begin{equation}
\xi _{transf}\left( _{\mathbf{Q}_{1}^{\prime }\ \ \mathbf{Q}_{1}}^{\mathbf{Q}%
_{2}^{\prime }\ \ \mathbf{Q}_{2}}\right) =(\mathcal{V}_{\mathbf{Q}%
_{1}^{\prime }}+\mathcal{V}_{\mathbf{Q}_{2}^{\prime }})\lambda \left( _{%
\mathbf{Q}_{1}^{\prime }\ \ \mathbf{Q}_{1}}^{\mathbf{Q}_{2}^{\prime }\ \ 
\mathbf{Q}_{2}}\right)  \tag{8.19}
\end{equation}%
While its contribution in the sum (8.1) definitely differs from zero, this
expression of $\xi _{transf}$ leads us to see it as a "transfer assisted
exchange" (see Fig. 5b). And indeed, it appears in the interaction
scattering $\xi $ or $\widetilde{\xi }$ with the minus sign as for any
exchange process. To better grasp this term, we can note that, although the
potential $S_{X}$ from which it originates, is a \textit{one-body} operator
in the Frenkel exciton subspace, it can lead to a scattering between \textit{%
two} excitons thanks to carrier exchanges allowed by the exciton composite
nature. Such a scattering $\xi _{transf}$, which does not exist for Wannier
excitons when electron-hole exchange, i.e., valence-conduction Coulomb
process, is neglected, is quite specific to Frenkel excitons: Indeed,
besides direct Coulomb processes which exist for both excitons, Frenkel
excitons in addition have indirect Coulomb processes which provide other
energy-like quantities from which this novel energy-like scattering can be
constructed, by mixing them with carrier exchanges.

It can be of interest to relate this $\xi _{transf}$ scattering to the novel
scattering we have identified between two polaritons \cite{12} - we called
"photon assisted exchange". For polaritons, the Pauli scattering induced by
carrier exchanges between the exciton components, is mixed with the Rabi
coupling between photon and exciton, which also is a relevant energy-like
quantity appearing in the one-body part of the polariton hamiltonian. As for
this photon assisted exchange, the strange asymmetry between "in" and "out"
states of assisted exchange scatterings is actually cured in real physical
processes in which energy is conserved since the scatterings with
interaction before or after the exchange are then equal, as shown in Section
IX.

\subsection{Exchange interaction scatterings}

The exchange Coulomb scatterings which are succession of carrier exchange
before or after carrier interaction, play a key role for Wannier excitons.
They are the dominant scatterings for excitons with close-to-zero momenta.
We are going to show that in the case of Frenkel excitons these exchange
interaction scatterings reduce to zero%
\begin{equation}
\widetilde{\xi }^{in}\left( _{\mathbf{Q}_{1}^{\prime }\ \ \mathbf{Q}_{1}}^{%
\mathbf{Q}_{2}^{\prime }\ \ \mathbf{Q}_{2}}\right) =\sum_{\mathbf{P}_{1}%
\mathbf{P}_{2}}\lambda \left( _{\mathbf{Q}_{1}^{\prime }\ \ \mathbf{P}_{1}}^{%
\mathbf{Q}_{2}^{\prime }\ \ \mathbf{P}_{2}}\right) \widetilde{\xi }\left( _{%
\mathbf{P}_{1}\ \ \mathbf{Q}_{1}}^{\mathbf{P}_{2}\ \ \mathbf{Q}_{2}}\right)
=0  \tag{8.20}
\end{equation}

By using Eq. (3.11) for the Pauli scattering, we readily find that the "in"
exchange scattering associated to direct Coulomb processes given in Eq.
(8.13) reduces to zero%
\begin{equation}
\xi _{coul}^{in}\left( _{\mathbf{Q}_{1}^{\prime }\ \ \mathbf{Q}_{1}}^{%
\mathbf{Q}_{2}^{\prime }\ \ \mathbf{Q}_{2}}\right) =\frac{1}{N_{s}^{2}}%
\delta _{\mathbf{Q}_{1}^{\prime }+\mathbf{Q}_{2}^{\prime },\mathbf{Q}_{1}+%
\mathbf{Q}_{2}}\sum_{\mathbf{Q}}\mathcal{W}_{\mathbf{Q}}=0  \tag{8.21}
\end{equation}%
due to Eq. (2.19), namely $\sum_{\mathbf{Q}}e^{i\mathbf{Q.\mathbf{R}}%
}=N_{s}\delta _{\mathbf{R},\mathbf{0}}$ since the $\mathbf{R=0}$ term are
also missing for the sums which enter $\mathcal{W}_{\mathbf{Q}}$ (see Eq.
(8.11)).

In the same way, the "in" exchange scattering associated to the transfer
part of the interaction given in Eq. (8.14) reduces to zero%
\begin{equation}
\xi _{transf}^{in}\left( _{\mathbf{Q}_{1}^{\prime }\ \ \mathbf{Q}_{1}}^{%
\mathbf{Q}_{2}^{\prime }\ \ \mathbf{Q}_{2}}\right) =\frac{1}{N_{s}^{2}}%
\delta _{\mathbf{Q}_{1}^{\prime }+\mathbf{Q}_{2}^{\prime },\mathbf{Q}_{1}+%
\mathbf{Q}_{2}}\sum_{\mathbf{Q}}(\mathcal{V}_{\mathbf{Q}_{1}+\mathbf{Q}}+%
\mathcal{V}_{\mathbf{Q}_{2}-\mathbf{Q}})=0  \tag{8.22}
\end{equation}%
since the $\mathbf{R=0}$ term is also missing from the sum entering $%
\mathcal{V}_{\mathbf{Q}}$ (see Eq. (2.15)).

We thus conclude that, as obtained from very general arguments at the
beginning of this section, the "in" exchange scattering for the first two
parts of $\xi \left( _{\mathbf{Q}_{1}^{\prime }\ \ \mathbf{Q}_{1}}^{\mathbf{Q%
}_{2}^{\prime }\ \ \mathbf{Q}_{2}}\right) $, namely $\xi _{coul}^{in}\left(
_{\mathbf{Q}_{1}^{\prime }\ \ \mathbf{Q}_{1}}^{\mathbf{Q}_{2}^{\prime }\ \ 
\mathbf{Q}_{2}}\right) $ and $\xi _{transf}^{in}\left( _{\mathbf{Q}%
_{1}^{\prime }\ \ \mathbf{Q}_{1}}^{\mathbf{Q}_{2}^{\prime }\ \ \mathbf{Q}%
_{2}}\right) $ reduce to zero while the third part $\xi _{neut}$ gives zero
when inserted in the relevant sum (see Eq. (8.16)). This physically means
that the energy-like part of the Frenkel exciton scattering is only
controlled by direct Coulomb processes \textit{and} by transfer assisted
exchange, a result not obvious at first.

\section{Frenkel exciton Hamiltonian in the two-exciton subspace}

\subsection{$H_{X}$ acting on two excitons}

In order to better grasp the interplay between the various scatterings
appearing in the preceding sections, let us calculate the matrix element of
the interacting exciton Hamiltonian in the two-exciton subspace. Using Eq.
(7.1), we find that $H_{X}$\ acting on two Frenkel excitons gives%
\begin{equation}
H_{X}B_{\mathbf{Q}_{1}}^{\dag }B_{\mathbf{Q}_{2}}^{\dag }\left\vert
v\right\rangle =\left( B_{\mathbf{Q}_{1}}^{\dag }H_{X}+E_{\mathbf{Q}_{1}}B_{%
\mathbf{Q}_{1}}^{\dag }+V_{\mathbf{Q}_{1}}^{\dag }\right) B_{\mathbf{Q}%
_{2}}^{\dag }\left\vert v\right\rangle   \tag{9.1}
\end{equation}%
Due to Eq. (8.1), this also reads%
\begin{equation*}
\left( H_{X}-E_{\mathbf{Q}_{1}}-E_{\mathbf{Q}_{2}}\right) B_{\mathbf{Q}%
_{1}}^{\dag }B_{\mathbf{Q}_{2}}^{\dag }\left\vert v\right\rangle 
\end{equation*}%
\begin{equation}
=V_{\mathbf{Q}_{1}}^{\dag }B_{\mathbf{Q}_{2}}^{\dag }\left\vert
v\right\rangle =\sum\limits_{\mathbf{P}_{1}\text{ }\mathbf{P}_{2}}\xi \left(
_{\mathbf{P}_{1}\ \mathbf{Q}_{1}}^{\mathbf{P}_{2}\ \mathbf{Q}_{2}}\right) B_{%
\mathbf{P}_{1}}^{\dagger }B_{\mathbf{P}_{2}}^{\dagger }\left\vert
v\right\rangle   \tag{9.2}
\end{equation}%
where $\xi $\ can be possibly replaced by $\widetilde{\xi }$ defined in Eq.
(8.18), due to Eq. (8.16).

\subsection{Matrix element of $H_{X}$ in the two-exciton subspace}

We now turn to the matrix element of the interacting exciton Hamiltonian $%
H_{X}$ in the two exciton subspace. With $H_{X}$ acting on the right, we find%
\begin{equation*}
\left\langle v\right\vert B_{\mathbf{Q}_{2}^{\prime }}B_{\mathbf{Q}%
_{1}^{\prime }}H_{X}B_{\mathbf{Q}_{1}}^{\dag }B_{\mathbf{Q}_{2}}^{\dag
}\left\vert v\right\rangle =\left( E_{\mathbf{Q}_{1}}+E_{\mathbf{Q}%
_{2}}\right) \left\langle v\right\vert B_{\mathbf{Q}_{2}^{\prime }}B_{%
\mathbf{Q}_{1}^{\prime }}B_{\mathbf{Q}_{1}}^{\dag }B_{\mathbf{Q}_{2}}^{\dag
}\left\vert v\right\rangle 
\end{equation*}%
\begin{equation}
+\sum_{\mathbf{P}_{1}\text{ }\mathbf{P}_{2}}\left\langle v\right\vert B_{%
\mathbf{Q}_{2}^{\prime }}B_{\mathbf{Q}_{1}^{\prime }}B_{\mathbf{P}%
_{1}}^{\dag }B_{\mathbf{P}_{2}}^{\dag }\left\vert v\right\rangle \widetilde{%
\xi }\left( _{\mathbf{P}_{1}\ \mathbf{Q}_{1}}^{\mathbf{P}_{2}\ \mathbf{Q}%
_{2}}\right)   \tag{9.3}
\end{equation}

If we now use the scalar product of two-exciton states given in Eq. (5.2),
the first term of Eq. (9.3) reduces to%
\begin{equation}
\left( E_{\mathbf{Q}_{1}}+E_{\mathbf{Q}_{2}}\right) \left( \delta _{\mathbf{Q%
}_{1}^{\prime }\mathbf{Q}_{1}}\delta _{\mathbf{Q}_{2}^{\prime }\mathbf{Q}%
_{2}}+\delta _{\mathbf{Q}_{1}^{\prime }\mathbf{Q}_{2}}\delta _{\mathbf{Q}%
_{2}^{\prime }\mathbf{Q}_{1}}-\frac{2}{N_{s}}\delta _{\mathbf{Q}_{1}^{\prime
}+\mathbf{Q}_{2}^{\prime },\mathbf{Q}_{1}+\mathbf{Q}_{2}}\right)  \tag{9.4}
\end{equation}

To calculate the second term, we use the same scalar product and take into
account that the exchange scatterings reduce to zero according to Eq.
(8.20). This second term then reads%
\begin{equation}
\widetilde{\xi }\left( _{\mathbf{Q}_{1}^{\prime }\ \ \mathbf{Q}_{1}}^{%
\mathbf{Q}_{2}^{\prime }\ \ \mathbf{Q}_{2}}\right) +\widetilde{\xi }\left( _{%
\mathbf{Q}_{2}^{\prime }\ \ \mathbf{Q}_{1}}^{\mathbf{Q}_{1}^{\prime }\ \ 
\mathbf{Q}_{2}}\right) =\frac{1}{N_{s}}\left( \mathcal{W}_{\mathbf{Q}%
_{1}^{\prime }-\mathbf{Q}_{1}}+\mathcal{W}_{\mathbf{Q}_{2}^{\prime }-\mathbf{%
Q}_{1}}-2\mathcal{V}_{\mathbf{Q}_{2}^{\prime }}-2\mathcal{V}_{\mathbf{Q}%
_{1}^{\prime }}\right) \delta _{\mathbf{Q}_{1}^{\prime }+\mathbf{Q}%
_{2}^{\prime },\mathbf{Q}_{1}+\mathbf{Q}_{2}}  \tag{9.5}
\end{equation}

\subsection{On the asymmetry of transfer assisted exchange scatterings}

In order to better accept the asymmetry between "in" and "out" excitons
appearing in the transfer assisted exchange scattering, let us now show that
these "in" and "out" scatterings are equal when energy is conserved. For
that, we use the above equations to calculate the same matrix element $%
\left\langle v\right\vert B_{\mathbf{Q}_{2}^{\prime }}B_{\mathbf{Q}%
_{1}^{\prime }}H_{X}B_{\mathbf{Q}_{1}}^{\dag }B_{\mathbf{Q}_{2}}^{\dag
}\left\vert v\right\rangle $ but through%
\begin{equation*}
\left\langle v\right\vert B_{\mathbf{Q}_{2}}B_{\mathbf{Q}_{1}}H_{X}B_{%
\mathbf{Q}_{1}^{\prime }}^{\dag }B_{\mathbf{Q}_{2}^{\prime }}^{\dag
}\left\vert v\right\rangle ^{\ast }
\end{equation*}%
\begin{equation*}
=\left( E_{\mathbf{Q}_{1}^{\prime }}^{\ast }+E_{\mathbf{Q}_{2}^{\prime
}}^{\ast }\right) \left( \delta _{\mathbf{Q}_{1}^{\prime }\mathbf{Q}%
_{1}}\delta _{\mathbf{Q}_{2}^{\prime }\mathbf{Q}_{2}}+\delta _{\mathbf{Q}%
_{1}^{\prime }\mathbf{Q}_{2}}\delta _{\mathbf{Q}_{2}^{\prime }\mathbf{Q}%
_{1}}-\frac{2}{N_{s}}\delta _{\mathbf{Q}_{1}^{\prime }+\mathbf{Q}%
_{2}^{\prime },\mathbf{Q}_{1}+\mathbf{Q}_{2}}\right) 
\end{equation*}
\begin{equation*}
+\frac{1}{N_{s}}\left( \mathcal{W}_{\mathbf{Q}_{1}-\mathbf{Q}_{1}^{\prime
}}^{\ast }+\mathcal{W}_{\mathbf{Q}_{2}-\mathbf{Q}_{2}^{\prime }}^{\ast }-2%
\mathcal{V}_{\mathbf{Q}_{2}}^{\ast }-2\mathcal{V}_{\mathbf{Q}_{1}}^{\ast
}\right) \delta _{\mathbf{Q}_{1}+\mathbf{Q}_{2},\mathbf{Q}_{1}^{\prime }+%
\mathbf{Q}_{2}^{\prime }}
\end{equation*}%
By noting that $E_{\mathbf{Q}}$ is real as well as $\mathcal{V}_{\mathbf{Q}}$
(see Eq. (2.15)), while $\mathcal{W}_{\mathbf{Q}}^{\ast }=\mathcal{W}_{%
\mathbf{Q}}=\mathcal{W}_{-\mathbf{Q}}$\ (see Eq. (8.11)), and by comparing
the two expressions of $\left\langle v\right\vert B_{\mathbf{Q}_{2}^{\prime
}}B_{\mathbf{Q}_{1}^{\prime }}H_{X}B_{\mathbf{Q}_{1}}^{\dag }B_{\mathbf{Q}%
_{2}}^{\dag }\left\vert v\right\rangle $, we readily see that%
\begin{equation*}
\left( \mathcal{V}_{\mathbf{Q}_{1}^{\prime }}+\mathcal{V}_{\mathbf{Q}%
_{2}^{\prime }}-\mathcal{V}_{\mathbf{Q}_{1}}-\mathcal{V}_{\mathbf{Q}%
_{2}}\right) \delta _{\mathbf{Q}_{1}^{\prime }+\mathbf{Q}_{2}^{\prime },%
\mathbf{Q}_{1}+\mathbf{Q}_{2}}
\end{equation*}%
\begin{equation}
=\left( E_{\mathbf{Q}_{1}^{\prime }}+E_{\mathbf{Q}_{2}^{\prime }}-E_{\mathbf{%
Q}_{1}}-E_{\mathbf{Q}_{2}}\right) \delta _{\mathbf{Q}_{1}^{\prime }+\mathbf{Q%
}_{2}^{\prime },\mathbf{Q}_{1}+\mathbf{Q}_{2}}  \tag{9.7}
\end{equation}%
which also reads%
\begin{equation*}
\xi _{transf}\left( _{\mathbf{Q}_{1}^{\prime }\ \ \mathbf{Q}_{1}}^{\mathbf{Q}%
_{2}^{\prime }\ \ \mathbf{Q}_{2}}\right) -\xi _{transf}\left( _{\mathbf{Q}%
_{1}\ \ \mathbf{Q}_{1}^{\prime }}^{\mathbf{Q}_{2}\ \ \mathbf{Q}_{2}^{\prime
}}\right) 
\end{equation*}%
\begin{equation}
=\left( E_{\mathbf{Q}_{1}^{\prime }}+E_{\mathbf{Q}_{2}^{\prime }}-E_{\mathbf{%
Q}_{1}}-E_{\mathbf{Q}_{2}}\right) \lambda \left( _{\mathbf{Q}_{1}^{\prime }\
\ \mathbf{Q}_{1}}^{\mathbf{Q}_{2}^{\prime }\ \ \mathbf{Q}_{2}}\right)  
\tag{9.8}
\end{equation}%
This proves that the asymmetry between "in" and "out" states in the transfer
term is removed for energy conserving processes which are the ones
controlling the physically relevant processes between two Frenkel excitons.
A similar result is found for the photon assisted exchange scattering of two
polaritons.

\subsection{Hamiltonian expectation value}

From Eqs. (5.2) and (9.3-5), we get the Hamiltonian expectation value for
two Frenkel excitons as%
\begin{equation*}
\frac{\left\langle v\right\vert B_{\mathbf{Q}_{1}}B_{\mathbf{Q}_{2}}H_{X}B_{%
\mathbf{Q}_{1}}^{\dag }B_{\mathbf{Q}_{2}}^{\dag }\left\vert v\right\rangle }{%
\left\langle v\right\vert B_{\mathbf{Q}_{1}}B_{\mathbf{Q}_{2}}B_{\mathbf{Q}%
_{1}}^{\dag }B_{\mathbf{Q}_{2}}^{\dag }\left\vert v\right\rangle }=E_{%
\mathbf{Q}_{1}}+E_{\mathbf{Q}_{2}}
\end{equation*}%
\begin{equation}
+\frac{1}{N_{s}}\frac{\mathcal{W}_{\mathbf{0}}+\mathcal{W}_{\mathbf{Q}_{2}-%
\mathbf{Q}_{1}}-2\mathcal{V}_{\mathbf{Q}_{1}}-2\mathcal{V}_{\mathbf{Q}_{2}}}{%
1+\delta _{\mathbf{Q}_{1}\text{ }\mathbf{Q}_{2}}-\frac{2}{N_{s}}}  \tag{9.9}
\end{equation}%
So that the Hamiltonian expectation value for two identical excitons $%
\mathbf{Q}_{1}=\mathbf{Q}_{2}=\mathbf{Q}$ reduces to%
\begin{equation}
\left\langle H_{X}\right\rangle _{2}=2E_{\mathbf{Q}}+\frac{\mathcal{W}_{%
\mathbf{0}}-2\mathcal{V}_{\mathbf{Q}}}{N_{s}-1}  \tag{9.10}
\end{equation}%
Its interaction part contains a $\mathcal{W}_{\mathbf{0}}$ contribution
which comes from direct Coulomb processes between sites. It also contains an
indirect "transfer" contribution induced by the possible fermion exchanges
between composite boson excitons. This term which comes from exchange
between \textit{two} excitons mixed with excitation transfer has a nature
quite different from the one of the direct Coulomb term $\mathcal{W}_{%
\mathbf{0}}$. As $E_{\mathbf{Q}}$, it depends on the momentum $\mathbf{Q}$
of the exciton considered. We see that the interaction part of the
Hamiltonian expectation value of two excitons reduces to zero in the large
sample limit, i.e., for $N_{s}\rightarrow \infty $. This is physically
reasonable since the interaction of just two excitons in a huge volume
cannot change their energy very much. However, as in the case of Wannier
excitons, this interaction part is going to appear with a $N(N-1)/2$\
prefactor when considering $N$ excitons instead of two, due to the number of
ways to choose 2 excitons among $N$. The precise calculation of the
Hamiltonian expectation value for $N$ Frenkel excitons is out of the scope
of the present paper.

\subsection{Comparison with elementary bosons}

Since many people are still tempted to replace excitons by elementary bosons
with a set of "appropriate" effective scatterings dressed by exchanges, let
us again show that, as for Wannier excitons, it is not possible to construct
such an effective bosonic Hamiltonian for Frenkel excitons, due to possible
carrier exchanges between these excitons.

If an effective Hamiltonian were to exist, it would read%
\begin{equation}
\overline{H}_{X}=\sum E_{\mathbf{P}}\overline{B}_{\mathbf{P}}^{\dag }%
\overline{B}_{\mathbf{P}}+\frac{1}{2}\sum \overline{V}\left( _{\mathbf{P}%
_{1}^{\prime }\ \mathbf{P}_{1}}^{\mathbf{P}_{2}^{\prime }\ \mathbf{P}%
_{2}}\right) \overline{B}_{\mathbf{P}_{1}^{\prime }}^{\dag }\overline{B}_{%
\mathbf{P}_{2}^{\prime }}^{\dag }\overline{B}_{\mathbf{P}_{2}}\overline{B}_{%
\mathbf{P}_{1}}  \tag{9.11}
\end{equation}%
with $\left[ \overline{B}_{\mathbf{P}^{\prime }},\overline{B}_{\mathbf{P}%
}^{\dag }\right] =\delta _{\mathbf{P}^{\prime }\mathbf{P}}$ for boson
excitons; so that $\overline{H}_{X}$ acting on two-exciton states would give%
\begin{equation*}
\overline{H}_{X}\overline{B}_{\mathbf{Q}_{1}}^{\dag }\overline{B}_{\mathbf{Q}%
_{2}}^{\dag }\left\vert v\right\rangle =(E_{\mathbf{Q}_{1}}+E_{\mathbf{Q}%
_{2}})\overline{B}_{\mathbf{Q}_{1}}^{\dag }\overline{B}_{\mathbf{Q}%
_{2}}^{\dag }\left\vert v\right\rangle 
\end{equation*}%
\begin{equation}
+\frac{1}{2}\sum \left( \overline{V}\left( _{\mathbf{Q}_{1}^{\prime }\ 
\mathbf{Q}_{1}}^{\mathbf{Q}_{2}^{\prime }\ \mathbf{Q}_{2}}\right) +\overline{%
V}\left( _{\mathbf{Q}_{1}^{\prime }\ \mathbf{Q}_{2}}^{\mathbf{Q}_{2}^{\prime
}\ \mathbf{Q}_{1}}\right) \right) \overline{B}_{\mathbf{Q}_{1}^{\prime
}}^{\dag }\overline{B}_{\mathbf{Q}_{2}^{\prime }}^{\dag }\left\vert
v\right\rangle   \tag{9.12}
\end{equation}%
To get the effective scatterings $\overline{V}\left( _{\mathbf{Q}%
_{1}^{\prime }\ \mathbf{Q}_{2}}^{\mathbf{Q}_{2}^{\prime }\ \mathbf{Q}%
_{1}}\right) $, we can think of comparing this result with $H_{X}$ acting on
two Frenkel excitons. By symmetrizing Eq. (9.2) for $B_{\mathbf{Q}%
_{1}^{\prime }}^{\dagger }B_{\mathbf{Q}_{2}^{\prime }}^{\dagger }=B_{\mathbf{%
Q}_{2}^{\prime }}^{\dagger }B_{\mathbf{Q}_{1}^{\prime }}^{\dagger }$, this
leads us to take $\overline{V}\left( _{\mathbf{Q}_{1}^{\prime }\ \mathbf{Q}%
_{1}}^{\mathbf{Q}_{2}^{\prime }\ \mathbf{Q}_{2}}\right) $ as $\xi \left( _{%
\mathbf{Q}_{1}^{\prime }\ \mathbf{Q}_{1}}^{\mathbf{Q}_{2}^{\prime }\ \mathbf{%
Q}_{2}}\right) $ or better as $\widetilde{\xi }\left( _{\mathbf{Q}%
_{1}^{\prime }\ \mathbf{Q}_{1}}^{\mathbf{Q}_{2}^{\prime }\ \mathbf{Q}%
_{2}}\right) $ in order to avoid the spurious constant term $(-2\varepsilon
_{0})$ in the effective scattering. However, this result is fully
unacceptable since the ($\mathcal{V}_{\mathbf{Q}_{1}^{\prime }}+\mathcal{V}_{%
\mathbf{Q}_{2}^{\prime }}$) part of $\xi _{transf}$ induces a
non-hermiticity in the effective Hamiltonian. Indeed, for $\overline{H}_{X}$
to be hermitian, we must have%
\begin{equation}
\overline{V}\left( _{\mathbf{P}_{1}^{\prime }\ \mathbf{P}_{1}}^{\mathbf{P}%
_{2}^{\prime }\ \mathbf{P}_{2}}\right) =\left[ \overline{V}\left( _{\mathbf{P%
}_{1}\ \mathbf{P}_{1}^{\prime }}^{\mathbf{P}_{2}\ \mathbf{P}_{2}^{\prime
}}\right) \right] ^{\ast }  \tag{9.13}
\end{equation}%
while $\xi _{transf}$\ does not have this property. Consequently, as for
Wannier excitons \cite{14}, it is not possible to construct an effective
bosonic Hamiltonian for Frenkel excitons with the structure given by Eq.
(9.11), due to the presence of exchange processes in the matrix element.
However, we would like to notice that in Refs. \cite%
{AgrGal,Agranovich,AgrTos} the fermionic Hamiltonian was represented exactly
as an \textit{infinite} series expansion in creation and destruction
operators for excitations on atomic sites and these operators obey bosonic
commutation rules. Note that, within such a scheme, it is rather complicated
technically to treat processes involving large number of excitons, as
explained in a more detail in Section XI.

\section{Discussion}

Since Frenkel excitons are made of highly localized electron-hole pairs, we
can be led to think that carrier exchanges between Frenkel excitons should
be strongly reduced; so that their exciton composite nature should have very
little consequences on their many-body physics. To check this basic idea, we
are going to make a precise comparison between Frenkel excitons and
elementary bosons.

The fact that these excitons are made with electrons and holes localized on
the same site, should also make them rather different from Wannier excitons
constructed on \textit{delocalized} conduction electrons and valence holes.
A precise comparison of the basic properties of Frenkel and Wannier excitons
is thus of interest to better grasp the relevant parameters which control
the many-body physics of these composite bosons.

\subsection{Comparison with elementary bosons}

Let us first compare Frenkel excitons with elementary bosons.

Frenkel excitons are made of indistinguishable electrons and holes. They
thus are composite particles. However, due to the fact that the wave
functions of these carriers are the ones of highly localized atomic states
on the same ion site, there is one way only to construct two excitons out of
two electron-hole pairs. This leads to make Frenkel excitons appear as
elementary particles. To support this idea, we can mention their closure
relation which has the same $(1/N!)$ prefactor as the one for elementary
quantum particles.

In spite of these arguments, Frenkel excitons differ from elementary bosons
by various means. A very important aspect of their composite nature is the
existence of Pauli scatterings $\lambda \left( _{\mathbf{Q}_{1}^{\prime }\ \ 
\mathbf{Q}_{1}}^{\mathbf{Q}_{2}^{\prime }\ \ \mathbf{Q}_{2}}\right) $
between two Frenkel excitons induced by carrier exchanges. These
dimensionless scatterings in particular appear in the scalar product of two
Frenkel exciton states, which does not reduce to a set of Kronecker symbols,
as for elementary bosons. The possible carrier exchanges between excitons
also produce a normalization factor for $N$ identical excitons which is not $%
N!$ as for elementary bosons but $N!F_{N}$ where $F_{N}$ is not a small
correction but decreases exponentially with $N$. Finally, when mixed with
the indirect Coulomb processes responsible for the Frenkel exciton
excitation transfer between sites, these dimensionless Pauli scatterings
give rise to an energy-like "transfer assisted exchange scatterings" which
cannot be properly included in a bosonic effective Hamiltonian for Frenkel
excitons without dropping its required hermiticity.

These Frenkel excitons thus are somewhat awkward because they partly behave
as elementary particles and partly as composite particles. This makes their
handling through intuitive arguments rather dangerous. This is why the hard
algebra based on four commutators proposed in the present many-body theory,
appears as highly valuable to overcome any possible uncertainty.

\subsection{Comparison with Wannier excitons}

Let us now compare Wannier and Frenkel excitons.

Wannier excitons are constructed on two fully delocalized particles, the
conduction electron and the valence hole. The problem is then to make out of
these delocalized fermions a bound exciton. This is done through \textit{%
intraband} Coulomb processes. In contrast, Frenkel excitons are constructed
on two fully localized particles, the first atomic excited state of a given
site and the absence of atomic ground state electron on the same site. The
problem then is to delocalize this highly bound excitation to produce a
delocalized excitation. This is done through \textit{interband} Coulomb
processes.

Wannier excitons are characterized by a center-of-mass momentum and a
relative motion index, the electrons and holes on which these excitons are
constructed being two fully delocalized particles. In contrast, Frenkel
excitons are only characterized by a center-of-mass momentum since their
electrons and holes are not free but forced to be on the same site. Due to
this unique degree of freedom, $N$ Frenkel excitons have the $1/N!$
prefactor in their closure relation characteristic for $N$ elementary
particles, while the closure relation for Wannier excitons has a prefactor $%
\left( 1/N!\right) ^{2}$ which is the signature of the two degrees of
freedom of these excitons.

Being both composite particles, they can interact by fermion exchanges
through $2\times 2$ Pauli scatterings. These fermion exchanges make the
scalar product of $N$ identical excitons differ from their elementary boson
value $N!$ by a factor $F_{N}$, which, for both excitons, decreases
exponentially with $N\eta $. The small dimensionless parameter $\eta $\
associated to density $N/L^{D}$ which controls these fermion exchanges,
appears as $N\left( a_{x}/L\right) ^{D}$ for Wannier excitons, while it
appears as $N/N_{s}$\ for Frenkel excitons. This proves that the key
parameter for the many-body physics of excitons is not linked to the
extension of the electron-hole relative motion wave function as first
thought \ - otherwise $\eta $ would reduce to zero in the case of Frenkel
excitons. In contrast, it must be physically understood as the number
excitons at hand divided by the total number of excitons the sample can
accommodate, this number being the number of ion sites $N_{s}$ in the case
of Frenkel excitons while it is the ratio $\left( L/a_{x}\right) ^{D}$ of
the sample volume to the exciton volume in the case of Wannier excitons.

The Pauli scatterings for Wannier and Frenkel excitons are represented by
the same Shiva diagram for carrier exchange between two excitons. This
diagram, as well as the ones for carrier exchange between more than two
excitons, can be decomposed in terms of $2\times 2$ Pauli scatterings and
calculated along the same intuitive line as the one we have established for
Wannier excitons. The only "little" problem arises with two consecutive
exchanges: Instead of getting an identity, as visually seen, two consecutive
exchanges between two Frenkel excitons is found to reduce to an exchange. As
a direct consequence, it is not possible to rewrite two Frenkel excitons as
a sum of two other Frenkel excitons, as in the case of Wannier excitons.
This awkward result on two consecutive exchanges puts some shade on the very
visual handling of Shiva diagrams in the case of Frenkel excitons. This is
why it seems wise to support all\ visual results on Shiva diagrams by a hard
algebra of the Frenkel exciton matrix element they represent, using the four
key commutators of the present many-body theory.

The possible difficulties with Frenkel exciton Shiva diagrams, when compared
with Wannier excitons, can be traced back to the fact that the states on
which Frenkel excitons are constructed, do not form a clean complete basis
for \textit{all} electron-hole pair states due to the tight binding
approximation on which these excitons are based. This is why, when
summations over intermediate states have to be performed, they not always
give the correct results. To predict when these results are incorrect is
however not so easy. Thanks to the full proof procedure we have constructed,
it is however possible to reach the correct answer through a rather simple
algebra based on commutator manipulations.

Besides fermion exchanges, Frenkel and Wannier excitons also fell each other
through Coulomb interaction. In addition to direct processes between
carriers, indirect processes are also kept for Frenkel excitons since, in
the tight-binding limit for electron and hole states, these are the only
processes allowing to delocalize the excitation, i.e., to produce the
exciton. Although appearing as a one-body operator in the exciton subspace,
these indirect "transfer" processes give rise to a novel $2\times 2$
energy-like scattering, when mixed with carrier exchange. These "transfer
assisted exchange" scatterings which do not exist for Wannier excitons, are
specific to the Frenkel exciton many-body physics. They in particular appear
in the matrix element of the Hamiltonian in the two exciton subspace as well
as in the Hamiltonian expectation value.

\section{State of the art}

Some many-body properties of Frenkel excitons have already been studied in
the past, see Refs. \cite{AgrGal,Agranovich,AgrTos,Davydov,CherMuk,Mukamel}.

(i) Agranovich and coworkers \cite{AgrGal,Agranovich,AgrTos}, as well as
Davydov in Ref. \cite{Davydov}, start with a system Hamiltonian written in
terms of creation operators $a_{0n}^{\dagger }$ and $a_{1n}^{\dagger }$ for
ground and first excited states of isolated molecules on site $n$ which is
quite similar to our Hamiltonian with $a_{0n}$ essentially replaced by $%
b_{n}^{\dagger }$. Their fermionic Hamiltonian is then rewritten in terms of
excitation operators $B_{n}^{\dagger }$, i.e., products like $%
a_{1n}^{\dagger }a_{0n}$. To do so, it is invoked \cite{AgrGal} that the sum
of the occupation numbers for the ground and first excited states of each
molecule is equal to 1. This condition between scalars is then replaced by
the same condition between operators, as seen \ in Eq. (4.9) of Ref. \cite%
{AgrGal}, namely%
\begin{equation}
\widehat{N}_{0n}=1-\widehat{N}_{1n}  \tag{11.1}
\end{equation}%
where $\widehat{N}_{fn}=a_{fn}^{\dagger }a_{fn}$ is the occupation number
operator of the state $f=(0,1)$ in the site $n$. Eq. (11.1) is valid only
when applied to states in which the sum of the occupation numbers for the
ground and first excited states of each molecule is equal to 1. This
equation (11.1), when implemented to the theory significantly facilitates a
final compact form for the Hamiltonian. As seen from Refs. \cite%
{AgrGal,Agranovich}, this procedure has however been used for two-exciton
problems only. The generalization \cite{AgrTos} of this approach to an
arbitrary number of excitons turns out to be rather complicated technically,
in contrast to our method which is quite straightforward for any number of
Frenkel excitons, thanks to Eqs. (3.12, 13).

(ii) Mukamel and coworkers \cite{CherMuk,Mukamel,Muk1,Muk2} have followed a
completely different approach based on the idea that, for problems dealing
with a fixed number $N$ of electron-hole pairs, it is always possible to
write the system Hamiltonian $H$ as well as the commutators between
electron-hole operators $\left[ B_{m},B_{n}^{\dagger }\right] $ \textit{%
exactly}, in terms of \textit{infinite series} of electron-hole operators $%
B_{n_{1}}^{\dagger }...B_{n_{p}}^{\dagger }B_{m_{1}}...B_{m_{p}}$, namely%
\begin{equation*}
H=\sum_{p=1}^{\infty }H^{(p)}
\end{equation*}%
\begin{equation}
H^{(p)}=\sum_{\left\{ n\right\} \left\{ m\right\} }h^{(p)}(n_{1}\ldots
n_{p};m_{1}\ldots m_{p})B_{n_{1}}^{\dagger }...B_{n_{p}}^{\dagger
}B_{m_{1}}...B_{m_{p}}  \tag{11.2}
\end{equation}%
and similarly for $\left[ B_{m},B_{n}^{\dagger }\right] $. In order to
describe a system of $N$ interacting composite excitons, we then have to
keep terms in these series for the Hamiltonian and the pair commutator,
which contain products up to the $p=N$ and $p=N-1$, respectively. The
constant prefactors in front of all these operators are obtained through the
projection of the initial fermionic Hamiltonian and the commutator $\left[
B_{m},B_{n}^{\dagger }\right] $ on the $N$-pair subspace. This method which
is formally exact, turns out to be technically quite heavy since the series
become more and more complicated with the increase of $p$. Actually, the
authors of Refs. \cite{CherMuk,Mukamel,Muk1,Muk2} have derived these series
for two-exciton problems only, and they have used them to calculate
successfully the third order susceptibility $\chi ^{(3)}$.

Very recently, we have reconsidered the idea of using infinite series of
operators to describe interacting composite excitons \cite{Monique-seria}.
Instead of using free pair operators, like Mukamel and coworkers, we have
used correlated pair operators, i.e., excitons which are the relevant
operators for such problems. By making use of the exciton closure relation,
we have previously derived, we have been able to write the two infinite
series explicitly, through rather light calculations. These series read in
terms of Pauli and interaction scatterings. However, the proper handling of
these two infinite series for $N$-body problems turn out to be far more
cumbersome than the many-body theories for Wannier or Frenkel excitons we
have constructed, based on just four commutators.

\section{Conclusion}

In this paper, we propose a new many-body theory for Frenkel excitons in
which the composite nature of these particles is treated exactly, excitons
being never bosonized. Starting from the expression of the Frenkel exciton
creation operator $B_{\mathbf{Q}}^{\dag }$ written in terms of electron and
hole creation operators, we derive the commutation rules for these Frenkel
excitons. They clearly differ from the ones of elementary bosons due to
carrier exchanges induced by the exciton composite nature. These carrier
exchanges give rise to "Pauli scatterings" between excitons, similar to the
one we found for Wannier excitons. They are represented by the same Shiva
diagrams calculated along the same line. Due to the fact that Frenkel
excitons are only characterized by a center-of-mass momentum $\mathbf{Q}$,
their Pauli scatterings take an extremely simple form%
\begin{equation}
\lambda \left( _{\mathbf{Q}_{1}^{\prime }\ \mathbf{Q}_{1}}^{\mathbf{Q}%
_{2}^{\prime }\ \mathbf{Q}_{2}}\right) =\frac{1}{N_{s}}\delta _{\mathbf{Q}%
_{1}^{\prime }+\mathbf{Q}_{2}^{\prime },\text{ }\mathbf{Q}_{1}+\mathbf{Q}%
_{2}}  \tag{12.1}
\end{equation}%
where $N_{s}$ is the number of ion sites, the many-body physics of $N$
Frenkel excitons being controlled by the small dimensionless parameter 
\begin{equation}
\eta =N/N_{s}  \tag{12.2}
\end{equation}%
This parameter is the analog of parameter $\eta =N\left( a_{x}/L\right) ^{D}$
for Wannier excitons since $N_{s}$ like $\left( L/a_{x}\right) ^{D}$ is the
maximum number of excitons the sample can accommodate.

As Frenkel excitons are constructed on highly localized electron-hole pairs,
we could expect them to behave as elementary bosons. While their closure
relation is indeed the one of elementary bosons - with a prefactor $1/N!$
while the one for Wannier excitons has a prefactor $\left( 1/N!\right) ^{2}$
- we find that they definitely behave as composite bosons through many other
properties. In particular, the normalization factor of $N$ Frenkel excitons $%
\left\langle v\right\vert B_{\mathbf{Q}}^{N}B_{\mathbf{Q}}^{\dag
N}\left\vert v\right\rangle =N!F_{N}$ differs from its elementary boson
value $N!$ by a factor%
\begin{equation}
F_{N}=\frac{N_{s}!}{(N_{s}-N)!N_{s}^{N}}  \tag{12.3}
\end{equation}%
which, as for Wannier excitons, decreases exponentially with $N\eta $.

In addition to carrier exchanges, Frenkel excitons also interact through
Coulomb processes. The interaction scattering they produce, are derived, as
for Wannier excitons, through a set of creation potentials. The part of the
interaction scattering coming from direct Coulomb processes between sites
depends on the momentum transfer as%
\begin{equation}
\xi _{coul}\left( _{\mathbf{Q}_{1}+\mathbf{Q}\ \ \mathbf{Q}_{1}}^{\mathbf{Q}%
_{2}-\mathbf{Q}^{\prime }\ \mathbf{Q}_{2}}\right) =\frac{1}{N_{s}}\delta _{%
\mathbf{\mathbf{Q}}\text{ }\mathbf{\mathbf{Q}}^{\prime }}\mathcal{W}_{%
\mathbf{Q}}  \tag{12.4}
\end{equation}

\begin{equation}
\mathcal{W}_{\mathbf{Q}}=\sum_{\mathbf{R}\neq \mathbf{0}}e^{-i\mathbf{Q}.%
\mathbf{R}}\left[ V_{\mathbf{R}}\left( _{1\ 1}^{1\ 1}\right) +V_{\mathbf{R}%
}\left( _{0\ 0}^{0\ 0}\right) -V_{\mathbf{R}}\left( _{0\ 1}^{1\ 0}\right)
-V_{\mathbf{R}}\left( _{1\ 0}^{0\ 1}\right) \right]  \tag{12.5}
\end{equation}%
where $V_{\mathbf{R}}\left( _{\nu _{1}^{\prime }\ \nu _{1}}^{\nu
_{2}^{\prime }\ \nu _{2}}\right) $ is the matrix element of the Coulomb
potential between atomic sites at $\mathbf{R}$, one electron going from the
atomic state $\nu _{1}$ to $\nu _{1}^{\prime }$ while the other goes from $%
\nu _{2}$ to $\nu _{2}^{\prime }$ (see Eq. (2.6)).

Besides this rather standard interaction scattering, we have identified a
novel "transfer assisted exchange" scattering which has similarity with
"photon assisted exchange" we found between polaritons. It comes from the
indirect Coulomb processes responsible for the excitation transfer of
Frenkel excitons, the coupling between two excitons being made through the
Pauli scattering for carrier exchanges between these excitons. This
"transfer assisted exchange" scattering depends on the "out" momenta ($%
\mathbf{Q}_{1}^{\prime }$, $\mathbf{Q}_{2}^{\prime }$) but not on the
momentum transfer $(\mathbf{Q}_{1}^{\prime }-\mathbf{Q}_{1})$ since the
excitation transfer does not induce any momentum change. Its precise value
reads%
\begin{equation}
\xi _{transf}\left( _{\mathbf{Q}_{1}^{\prime }\ \ \mathbf{Q}_{1}}^{\mathbf{Q}%
_{2}^{\prime }\ \ \mathbf{Q}_{2}}\right) =(\mathcal{V}_{\mathbf{Q}%
_{1}^{\prime }}+\mathcal{V}_{\mathbf{Q}_{2}^{\prime }})\lambda \left( _{%
\mathbf{Q}_{1}^{\prime }\ \mathbf{Q}_{1}}^{\mathbf{Q}_{2}^{\prime }\ \mathbf{%
Q}_{2}}\right)   \tag{12.6}
\end{equation}%
where $\mathcal{V}_{\mathbf{Q}}$ is the $\mathbf{Q}$\ dependent part of the
exciton energy.%
\begin{equation}
\mathcal{V}_{\mathbf{Q}}=\sum_{\mathbf{R}\neq 0}e^{-i\mathbf{Q.R}}V_{\mathbf{%
R}}\left( _{1\ 0}^{0\ 1}\right)   \tag{12.7}
\end{equation}%
It is possible to show that, in this transfer assisted exchange, symmetry
between "in" and "out" states is restored for energy conserving processes
since we do have%
\begin{equation*}
\xi _{transf}\left( _{\mathbf{Q}_{1}^{\prime }\ \ \mathbf{Q}_{1}}^{\mathbf{Q}%
_{2}^{\prime }\ \ \mathbf{Q}_{2}}\right) -\xi _{transf}\left( _{\mathbf{Q}%
_{1}\ \ \mathbf{Q}_{1}^{\prime }}^{\mathbf{Q}_{2}\ \ \mathbf{Q}_{2}^{\prime
}}\right) 
\end{equation*}%
\begin{equation}
=\left( E_{\mathbf{Q}_{1}^{\prime }}+E_{\mathbf{Q}_{2}^{\prime }}-E_{\mathbf{%
Q}_{1}}-E_{\mathbf{Q}_{2}}\right) \lambda \left( _{\mathbf{Q}_{1}^{\prime }\
\ \mathbf{Q}_{1}}^{\mathbf{Q}_{2}^{\prime }\ \ \mathbf{Q}_{2}}\right)  
\tag{12.8}
\end{equation}

Due to this "transfer assisted exchange" scattering, here also it is not
possible to construct an effective bosonic Hamiltonian for Frenkel excitons
which is hermitian, as physically required, and which produces the same
matrix element as the one of the exact fermionic Hamiltonian.

Frenkel excitons turn out to be quite tricky bosons: they appear as
elementary in some cases but composite in many others. This is why it seems
hard to guess physical results like the scattering rate of Frenkel excitons,
the ground state energy of $N$ Frenkel excitons or the nonlinear
susceptibilities of materials having such excitons, without a precise
calculation of these quantities using the procedure developed in this paper.
These problems will be addressed in a near future.

\acknowledgments

W. V. P. is supported by the Ministry of Education of France, the Russian
Science Support Foundation, and the President of Russia program for young
scientists.

\textbf{Figure captions}

Fig 1. (a) Electron-hole pair eigenstate $a_{n_{e}}^{\dagger }b_{n_{h}}^{\dagger
}\left\vert v\right\rangle $ of $H_{eh}$\ defined in Eq. (2.3).

(b) Electron-hole pair eigenstate $a_{n}^{\dagger
}b_{n}^{\dagger}\left\vert v\right\rangle $ of $H_{pair}$\ defined
in Eq. (2.4). 

(c) Indirect Coulomb potential $V_{trans}$ as
defined in Eq. (2.10) which allows to transfer the excitation from
site $n_{2}$ to site $n_{1}$. 

(d, e, h) Direct Coulomb potentials
$V_{eh}$, $V_{ee}$ and $V_{hh}$ as defined in Eqs. (2.26-28).

Fig 2. (a) Shiva diagram for the Pauli scattering
$\lambda \left( _{\mathbf{P}^{\prime }\
\mathbf{P}}^{\mathbf{Q}^{\prime }\ \mathbf{Q}}\right)$ of two "in"
Frenkel excitons ($\mathbf{P}$, $\mathbf{Q}$) toward two "out"
excitons ($\mathbf{P}^{\prime }$, $\mathbf{Q}^{\prime }$)\ as
defined in Eqs. (3.11) and (4.2, 4.4). 

(b) Shiva diagram for
carrier exchange of three Frenkel excitons defined in Eq. (4.5)
and its decomposition in $2\times 2$ Pauli scatterings as given in
Eq. (4.6). 

(c) Same as (b) for four excitons. 

(d) Two consecutive
carrier exchanges should reduce to an identity as seen from Fig.
2(e), while two consecutive Pauli scatterings do not reduce to
this identity, as seen from Eq. (4.7).

Fig 3. Shiva diagram for the scalar product of two
Frenkel exciton states, as given in Eq. (5.2).

Fig 4. Shiva diagram for the scalar product of
three Frenkel exciton states as given in Eq. (5.5).

Fig 5. (a) Part of the interaction scattering $\xi _{coul}\left( _{\mathbf{Q}%
_{1}^{\prime }\ \ \mathbf{Q}_{1}}^{\mathbf{Q}_{2}^{\prime }\ \ \mathbf{Q}%
_{2}}\right) $ associated to direct Coulomb processes between
electrons and holes of different sites defined in Eq. (8.13). 

(b)
Part of the interaction scattering $\xi _{trans}\left(
_{\mathbf{Q}_{1}^{\prime }\ \
\mathbf{Q}_{1}}^{\mathbf{Q}_{2}^{\prime }\ \ \mathbf{Q}
_{2}}\right)$ associated to the indirect Coulomb processes
insuring the excitation transfer defined in Eq. (8.14). This
conceptually new scattering can be seen as a "transfer assisted
exchange", similar to the "photon assisted exchange" scattering we
have identified between two polaritons.


\begin{references}

\bibitem{Wannier}G. H. Wannier, Phys. Rev. {\bf 52}, 191 (1937).

\bibitem{Frenkel}J. Frenkel, Phys. Rev. {\bf 37}, 17  (1931).

\bibitem{3}For a short review, see M. Combescot and O. Betbeder-Matibet, Solid State Comm.
{\bf 134}, 11 (2005).

\bibitem{4}For a longer review, see M. Combescot, O. Betbeder-Matibet, and F. Dubin,
Physics Reports {\bf 463}, 215 (2008).

\bibitem{Monique-Odile}M. Combescot and O. Betbeder-Matibet, Eur. Phys. J. B
{\bf 55}, 63 (2007).

\bibitem{a}M. Combescot and  O. Betbeder-Matibet, Phys. Rev. B
{\bf 74}, 125316 (2006).

\bibitem{a1}M. Combescot and  O. Betbeder-Matibet, cond-mat/0802.0435.

\bibitem{b}M. Combescot and  O. Betbeder-Matibet, Solid Stat. Com.
{\bf 132}, 129 (2004).

\bibitem{c}M. Combescot,  O. Betbeder-Matibet, and  V. Voliotis,
Europhys. Lett. {\bf 74}, 868 (2006).

\bibitem{d}M. Combescot,  O. Betbeder-Matibet, and  R. Combescot,
Phys. Rev. Lett. {\bf 99}, 176403 (2007).

\bibitem{paper1}M. Combescot and  W.  V. Pogosov, Phys. Rev. B
{\bf 77}, 085206 (2008).

\bibitem{7}M. Combescot and C. Tanguy, Europhys. Lett.
{\bf 55}, 390 (2001).

\bibitem{8}M. Combescot and O. Betbeder-Matibet, Europhys. Lett.
{\bf 58}, 87 (2002).

\bibitem{9}M. Combescot and O. Betbeder-Matibet, Eur. Phys. J. B
{\bf 27}, 505 (2002).

\bibitem{Monique-Tanguy}M. Combescot, X. Leyronas, and C. Tanguy, Eur. Phys. J. B
{\bf 31}, 17 (2003).

\bibitem{11}M. Combescot and O. Betbeder-Matibet, Phys. Rev. B
{\bf 72}, 193105 (2005).

\bibitem{Monique-Marc-Andre}M. Combescot and M. A. Dupertuis, Phys. Rev. B
{\bf 78}, 235303 (2008).

\bibitem{12}M. Combescot, M. A. Dupertuis, and O. Betbeder-Matibet, Europhys. Lett.
{\bf 75}, 17001 (2007).

\bibitem{13}O. Betbeder-Matibet and M. Combescot, Eur. Phys. J. B
{\bf 31}, 517 (2003).

\bibitem{14}M. Combescot, O. Betbeder-Matibet and R. Combescot, Phys. Rev. B
{\bf 75}, 114305 (2007).

\bibitem{AgrGal}V. M. Agranovich and M. D. Galanin, Electronic Excitation
Energy Transfer in Condensed Matter (North Holland, Amsterdam, 1982).

\bibitem{Agranovich}M. Hoffmann, K. Schmidt, T. Fritz, T. Hasche,
V.M. Agranovich, and K. Leo, Chem. Phys. {\bf 258}, 73 (2000).

\bibitem{AgrTos}V.  M. Agranovich and B. S. Toshich, Sov. Phys. JETP 26, 104 (1968).

\bibitem{Davydov}A. S. Davydov, Theory of Molecular Excitons (Plenum Press,
New York, 1971).

\bibitem{CherMuk}V. Chernyak and S. Mukamel, J. Opt. Soc. Am. B13, 1302 (1996).

\bibitem{Mukamel}S. Mukamel,
Principles of Nonlinear Optics and Spectroscopy, Oxford University Press (1995).

\bibitem{Muk1}V. Chernyak, S. Yokojima, T. Meier, and S. Mukamel,
Phys. Rev. B {\bf 58}, 4496  (1998).

\bibitem{Muk2}V. M. Axt and S. Mukamel,
Rev. Mod. Phys. {\bf 70}, 145  (1998).

\bibitem{Monique-seria}M. Combescot and O. Betbeder-Matibet, to appear in
Phys. Rev. B, cond-mat/0806.0435.

\end{references}
\end{document}